\documentclass[aip,jcp,numerical,reprint]{revtex4-1}
\usepackage[english]{babel}
\usepackage{microtype}
\usepackage{amsmath,amsthm,amssymb,amsfonts}
\usepackage{mathtools}
\usepackage{siunitx}
\usepackage{tabularx,graphicx}
\usepackage{booktabs}
\usepackage{xr}
\usepackage[breaklinks=true,colorlinks=true,linkcolor=blue,urlcolor=blue,citecolor=blue]{hyperref}

\usepackage{alphabeta}
\providecommand*{\Qoppa}{\text{\textQoppa}}

\usepackage[bordercolor=none,backgroundcolor=none,textsize=scriptsize]{todonotes}

\newcommand{\ddf}[1]{\mathrm{d} #1 \,}
\DeclarePairedDelimiter\avg{\langle}{\rangle}

\DeclareMathOperator{\tr}{tr}
\DeclareMathOperator\const{const.}
\DeclareMathOperator*{\argmax}{arg\,max}
\DeclareMathOperator*{\argmin}{arg\,min}
\newcommand{\e}{\mathrm{e}}

\newcommand{\NM}{\mathcal{N}_{M}}

\usepackage{mhchem}

\DeclareSIUnit\molar{\textsc{M}}

\newcommand{\var}{\operatorname{var}}

\begin{document}

\preprint{}

\title{Maximum likelihood estimates of diffusion coefficients from single-particle tracking experiments}

\author{Jakob T\'{o}mas Bullerjahn}
\affiliation{Department of Theoretical Biophysics, Max Planck Institute of Biophysics, 60438 Frankfurt am Main, Germany}
\author{Gerhard~Hummer}
\email{gerhard.hummer@biophys.mpg.de}
\affiliation{Department of Theoretical Biophysics, Max Planck Institute of Biophysics, 60438 Frankfurt am Main, Germany}
\affiliation{Institute of Biophysics, Goethe University Frankfurt, 60438 Frankfurt am Main, Germany}
\date{\today}

\begin{abstract}
Single-molecule localization microscopy allows practitioners to locate and track labeled molecules in biological systems.  When extracting diffusion coefficients from the resulting trajectories, it is common practice to perform a linear fit on mean-square-displacement curves.  However, this strategy is suboptimal and prone to errors.  Recently, it was shown that the increments between observed positions provide a good estimate for the diffusion coefficient, and their statistics are well-suited for likelihood-based analysis methods.  Here, we revisit the problem of extracting diffusion coefficients from single-particle tracking experiments subject to static and dynamic noise using the principle of maximum likelihood.  Taking advantage of an efficient real-space formulation, we extend the model to mixtures of subpopulations differing in their diffusion coefficients, which we estimate with the help of the expectation-maximization algorithm.  This formulation naturally leads to a probabilistic assignment of trajectories to subpopulations.  We employ the theory to analyze experimental tracking data that cannot be explained with a single diffusion coefficient.  We test how well a dataset conforms to the assumptions of a diffusion model and determine the optimal number of subpopulations with the help of a quality factor of known analytical distribution.  To facilitate use by practitioners, we provide a fast open-source implementation of the theory for the efficient analysis of multiple trajectories in arbitrary dimensions simultaneously.  
\end{abstract}


\maketitle

\section{Introduction}

Single-particle tracking methods are routinely used to monitor the erratic motion of labeled macromolecules in their native environment, such as molecular motors moving along the cytoskeletal network~\cite{GellesSchnapp1988, YildizForkey2003}, transcription factors binding to DNA~\cite{GebhardtSuter2013}, or receptor proteins diffusing in cell membranes~\cite{SaxtonJacobson1997}.  The mode of transport and associated transport coefficients are inferred from the measured trajectories using microscopic models mimicking the globally observed behavior, which can range from ordinary to anomalous and confined diffusion, as well as active transport~\cite{LeviGratton2007}.  

The simplest mode of transport in a dense fluid medium is free diffusion, which is fully characterized by a diffusion coefficient $D$.  However, single-particle tracking experiments are plagued by static localization errors and dynamic motion blur, which have to be properly accounted for when estimating diffusion coefficients.  Static errors originate from various noise sources in the experimental setup, such as spatial resolution of the instrument, and noise in the detection and processing electronics~\cite{Bobroff1986}, and are commonly modeled via additive Gaussian noise~\cite{MartinForster2002, SavinDoyle2005, Berglund2010}.  Motion blur arises from the camera's finite frame integration time, during which the particle position is smeared out.  This effect depends on the illumination profile of the shutter, which in most cases is uniform~\cite{SavinDoyle2005}, but can, in general, be represented by any non-negative function $s(t)$ that integrates to unity over the frame integration time~\cite{Berglund2010}.  

Traditionally, the estimation of diffusion coefficients has relied on the fact that the slope of the mean squared displacement~(MSD) is directly proportional to $D$~\cite{QianSheetz1991}.  This procedure, however, has some serious drawbacks.  For instance, the most common estimator for the MSD of finite time series introduces correlations between the MSD values at different time lags, which, in turn, cause the estimate to suffer if too many MSD values are used for the fit~\cite{WieserSchutz2008, Michalet2010}.  To remedy these shortcomings, one can either consider the optimal number of MSD values for the fit~\cite{Michalet2010}, or explicitly account for the above-mentioned correlations in the fit procedure~\cite{Bullerjahnvon-Bulow2020}.  Departing from the idea that the MSD is needed to determine $D$, one arrives at the highly efficient covariance-based estimator~(CVE)~\cite{VestergaardBlainey2014}, which is based on the much simpler statistics of the position increments between two observations.  Even before, Berglund~\cite{Berglund2010} had already used the sparse covariance matrix of the increments to construct an approximate maximum likelihood estimator~(MLE) operating in discrete Fourier space.  This estimator asymptotically approaches the exact MLE in the limit of infinitely long trajectories.  It was later shown~\cite{MichaletBerglund2012} that replacing the Fourier transform with a discrete sine transform results in an orthogonal basis in which the MLE of Ref.~\onlinecite{Berglund2010} is exact.  

Despite the fact that the CVE is orders of magnitude faster than likelihood-based methods~\cite{VestergaardBlainey2014}, the latter can still be beneficial, \emph{e.g.}, to incorporate prior knowledge of the parameters in a Bayesian manner or to globally analyze trajectories of different lengths.  The latter issue is common in single-particle tracking, because individual molecules stochastically disappear as a result of photobleaching.  Importantly, MLEs are minimally affected by blinking events and other interruptions of the trajectory recording, in the sense that the estimates and their uncertainty depend almost exclusively on the number of observed particle positions, and not the number of trajectories. We note, though, that the CVE can also overcome missing positions in the trajectory by lumping together increments of the resulting trajectory segments~\cite{Vestergaard2016}.  However, correlations in longer trajectories may reveal deviations from regular diffusion.  

Here, we revisit the problem of minimizing the negative log-likelihood, but instead of transforming the data we work in real space, and exploit the symmetries of the increment covariance matrix to develop a fast and reliable numerical solution scheme.  We develop MLEs of the mean-squared positional uncertainty $a^{2}$ in single-particle tracking and of the diffusional spread $\sigma^{2}$ during a single time step $\Delta t$,
\begin{equation}
\sigma^{2} = 2 D \Delta t \, ,
\end{equation}
from which one obtains the MLE of the diffusion coefficient as $D = \sigma^{2} / (2 \Delta t)$, irrespective of the dimension $d$.  The MLEs can also be used to analyze molecular dynamics simulation trajectories by setting the motion-blur coefficient to $B=0$.  The coefficient $a^2$ then accounts for fast non-diffusional spread in the particle position.  

We extend the likelihood formulation to a mixture model, which assumes that the trajectories to be analyzed originate from different subpopulations, each characterized by a distinct diffusion coefficient.  In comparison to some established models, such as those that account for diffusion in inhomogeneous media~\cite{Hummer2005, MassonCasanova2009, Kleinhans2012, El-BeheiryDahan2015, HozeHolcman2015, KrogLomholt2017, LaurentFloderer2020, FrishmanRonceray2020} or those allowing for multiple diffusive regimes within a single trajectory~\cite{PerssonLinden2013, OttShai2013, MonnierBarry2015, LindenElf2018, FalcaoCoombs2020}, our approach may seem somewhat restrictive.   However, what it lacks in generality it makes up with speed, efficiency, and compactness.  Furthermore, we provide rigorous statistical tests to assess whether the data comply with the theory assumptions or if more demanding models are needed to explain the data.  

The paper is structured as follows.  In Sec.~\ref{sec:microscopic-model}, we review the minimal stochastic model of Ref.~\onlinecite{Berglund2010} for diffusive trajectories smeared out by static localization noise and dynamic motion blur.  Section~\ref{sec:mle} is dedicated to the negative log-likelihood of the trajectory increments and the efficient $\mathcal{O}(N)$-algorithm used to numerically minimize it.  To test the assumptions of the underlying diffusion model, we introduce a quality factor in Sec.~\ref{sec:quality-factor} that can either be inspected visually or via a non-parametric test, such as the Kolmogorov-Smirnov test or Kuiper's test.  In Sec.~\ref{sec:em-algorithm}, we generalize the likelihood function to a mixture of subpopulations, each with a distinct diffusion coefficient.  We minimize this joint likelihood with an expectation-maximization (EM) algorithm.  A novel selection criterion for the optimal number of subpopulations, based on quality factor statistics, is introduced and tested on synthetic data in Sec.~\ref{sec:selection-criterion}.  Section~\ref{sec:simulation-data} explores ways to determine whether a small sample of short trajectories is governed by diffusive dynamics or not.  In Sec.~\ref{sec:experimental-data}, we use the mixture MLEs to analyze single-molecule tracking experiments~\cite{HarwardtYoung2017} reporting on the diffusive dynamics of human MET receptor tyrosine kinase in live cells.  An outlook on possibilities to incorporate elements of Bayesian analysis in the theory is presented in Sec.~\ref{sec:bayesian-inference}.  Our findings are summarized in Sec.~\ref{sec:conclusions} and implemented in a data analysis package written in Julia~\cite{JuliaScript}.  Detailed derivations and background information on the theory can be found in the Appendix.

\section{Theory}\label{sec:theory}

\subsection{Microscopic model}\label{sec:microscopic-model}

A detailed stochastic model capturing the effective dynamics observed in single-molecule tracking experiments was introduced by Savin and Doyle~\cite{SavinDoyle2005} for a uniform illumination profile, and later generalized to arbitrary profiles by Berglund~\cite{Berglund2010}.  In what follows, we review the model by Berglund, on which we then build the MLE in Sec.~\ref{sec:mle}.  

Consider a freely diffusing particle in one dimension with diffusion coefficient $D$, whose dynamics is described with a time-continuous Wiener process $Y$, satisfying
\begin{equation}\label{eq:Y-process}
\dot{Y}(t) = \frac{\sigma}{\sqrt{\Delta t}} \xi(t) \, .  
\end{equation}
Here, a dot indicates the time derivative and $\xi(t)$ denotes Gaussian white noise fully characterized by $\avg{\xi(t)} = 0$ and $\avg{\xi(t) \xi(t')} = \delta (t-t')$ with $\delta(t)$ being the Dirac $\delta$-function.  The particle motion is captured by a camera with frame integration time $\Delta t$, during which the camera shutter may be fully or partially open.  This gives rise to the shutter function $s(t) \geq 0$ that smears out the particle position over the integration time $\Delta t$, resulting in the process
\begin{gather}\label{eq:Z-process}
Z_{i} = \int_{0}^{\Delta t} \ddf{\tau} s (\tau) Y \big( \tau + i \Delta t \big) \, ,
\\
\notag
\int_{0}^{\Delta t} \ddf{\tau} s(\tau) = 1 \, ,
\end{gather}
for $i = 1,2,\dots$.  Furthermore, we assume that each frame is affected by additional measurement noise, which we model as Gaussian with variance $\smash{a^{2} / 2}$.  The observed motion of the particle is therefore described by the following stochastic process,
\begin{equation}\label{eq:X-process}
X_{i} = Z_{i} + \frac{a}{\sqrt{2}} R_{i} \, ,
\end{equation}
where $R$ is a normal distributed random variable with $\avg{R_{i}} = 0$ and $\avg{R_{i} R_{j}} = \delta_{i,j}$, and $\delta_{i,j}$ is the Kronecker delta that evaluates to one if $i = j$ and zero otherwise.  Note that due to linearity, the Gaussian nature of $\xi$ and $R$ is inherited by the processes $X$, $Y$ and $Z$.  

According to Eqs.~\eqref{eq:Y-process}--\eqref{eq:X-process}, the observed mean-squared displacement (MSD) is given by
\begin{equation}\label{eq:msd}
\avg{(X_{i} - X_{0})^{2}} =  a^{2} + \sigma^{2} ( i - 2 B ) \, ,
\end{equation}
which is, in comparison to $\avg{[Y(i \Delta t) - Y(0)]^{2}} = i \sigma^{2}$, corrupted by the static error $a^{2}$ and the dynamic error $2 \sigma^{2} B$.  The latter is characterized by the motion blur coefficient
\begin{equation}
B = \frac{1}{\Delta t} \int_{0}^{\Delta t} \ddf{\tau} S(\tau) [ 1 - S(\tau) ] \, ,
\end{equation}
where $S(t)$ gives the relative amount of exposure up to the time $t$, namely
\begin{equation*}
S(t) = \int_{0}^{t} \ddf{\tau} s(\tau) \sim 
\begin{cases}
0 \, , & t = 0
\\
1 \, , & t = \Delta t
\end{cases} \, .  
\end{equation*}
For uniform illumination, we have $s(t) = \Delta t^{-1}$, which results in a motion blur coefficient of $B = 1/6$.  Generally $0 \leq B \leq 1/4$ must hold.  In the case of (near-)perfect time resolution, \emph{e.g.}, when analyzing molecular dynamics trajectories, one has $B=0$.

\subsection{Maximum likelihood estimation}\label{sec:mle}

The probability to observe a one-dimensional time series $\smash{\vec{X}} = \smash{( X_{0}, X_{1}, X_{2}, \dots, X_{N} )^{T}}$ is given by the Gaussian joint probability distribution function $p(\vec{X}) \propto \exp ( - [\vec{X} - \avg{\vec{X}}]^{T} \mathbf{\Sigma}_{X}^{-1} [\vec{X} - \avg{\vec{X}}] / 2)$, which can be reinterpreted as a likelihood function for the parameters $a^{2}$ and $\sigma^{2}$.  These enter the dense covariance matrix $\smash{\mathbf{\Sigma}_{X}}$ whose inversion requires $\mathcal{O}(N^{3})$ operations, and therefore makes the variation of $a^{2}$ and $\sigma^{2}$, when maximizing the likelihood, computationally expensive.  

The likelihood function can be expressed more economically in terms of the position increments $\Delta_{i} = X_{i} - X_{i-1}$ with $i = 1,2, \dots, N$.  Like the $X_{i}$ they are also Gaussian distributed, their expectation values read $\avg{\Delta_{i}} = 0$, and the elements of the corresponding covariance matrix $\mathbf{\Sigma}$ are given by
\begin{align}
\Sigma_{i,j}
\label{eq:covariance-matrix}
& = a^{2} \Sigma'_{i,j} + \sigma^{2} \Sigma''_{i,j} \, ,
\\
\label{eq:sigma-prime}
\Sigma'_{i,j} & = \delta_{i,j} - \frac{\delta_{i-1,j}}{2} - \frac{\delta_{i,j-1}}{2} \, ,
\\
\label{eq:sigma-double-prime}
\Sigma''_{i,j} & = (1 - 2 B) \delta_{i,j} + B (\delta_{i-1,j} + \delta_{i,j-1}) \, ,
\end{align}
\emph{i.e.}, $\mathbf{\Sigma}$ is a linear combination of the constant tridiagonal matrices $\mathbf{\Sigma'}$ and $\mathbf{\Sigma''}$.  The likelihood of observing the increments $\smash{\vec{\Delta}} = \smash{(\Delta_{1}, \Delta_{2}, \dots, \Delta_{N})^{T}}$ in one dimension thus has the form
\begin{equation}\label{eq:pdf}
\ell(\vec{\Delta} \mid a^{2}, \sigma^{2}) = \sqrt{\frac{1}{(2 \pi)^{N} \vert \mathbf{\Sigma} \vert}} \exp \bigg( - \frac{1}{2} \vec{\Delta}^{T} \mathbf{\Sigma}^{-1} \vec{\Delta} \bigg) \, ,
\end{equation}
which results in the following negative log-likelihood (up to a negligible constant),
\begin{equation}\label{eq:log-likelihood}
\mathcal{L} (\vec{\Delta} \mid a^{2}, \sigma^{2}) = \frac{1}{2} \vec{\Delta}^{T} \mathbf{\Sigma}^{-1} \vec{\Delta} + \frac{1}{2} \ln (\vert \mathbf{\Sigma} \vert) \, .  
\end{equation}
Because $\mathbf{\Sigma}$ is tridiagonal, Eq.~\eqref{eq:log-likelihood} can be evaluated in an efficient manner.  For example, the Thomas algorithm~\cite{QuarteroniSacco2000} and a three-term recurrence relation~\cite{El-Mikkawy2004} can be used to calculate $\smash{\mathbf{\Sigma}^{-1} \vec{\Delta}}$ and $\vert \mathbf{\Sigma} \vert$, respectively, in $\mathcal{O}(N)$ operations.  Yet, due to the fact that the covariance matrix here is also a symmetric Toeplitz matrix, the recurrence relation can be solved analytically (see Appendix~\ref{app:recurrence-relation}), giving
\begin{align}\label{eq:general-determinant}
\ln (\vert \mathbf{\Sigma} \vert)
\notag
& = N \ln (\alpha) + (N+1) \ln \bigg( \frac{1 + q}{2} \bigg)
\\
& \mathrel{\phantom{=}} + \ln \bigg( 1 - \bigg[ \frac{1 - q}{1 + q} \bigg]^{N+1} \bigg) - \ln (q) \, .  
\end{align}
Here, $q = \smash{\sqrt{1 - 4 \beta^{2} / \alpha^{2}}} \geq 0$ with $\alpha = a^{2} + \sigma^{2} (1 - 2 B)$ and $\beta = - a^{2} / 2 + \sigma^{2} B$.  If $q \to 0$, which implies either $\sigma \to 0$ or $a \to 0$ with $B=1/4$, the expression above reduces to
\begin{equation}\label{eq:q=0-determinant}
\lim_{q \to 0} \ln (\vert \mathbf{\Sigma} \vert) = \ln(N+1) + N \ln (\alpha / 2) \, .  
\end{equation}
Equations~\eqref{eq:general-determinant} and~\eqref{eq:q=0-determinant} can be modified to account for non-symmetric tridiagonal Toeplitz matrices, as demonstrated in Ref.~\onlinecite{QiLiu2018}.  

A global analysis of $M$ $d$-dimensional trajectories of different lengths $N_{m} + 1$ is realized by minimizing
\begin{align}\label{eq:M-log-likelihood}
\notag
\mathcal{L}(\{ & \vec{\Delta}_{m,n} \}_{m=1,\dots,M}^{n=1,\dots,d} \mid a^{2}, \sigma^{2})
\\
& = \frac{1}{2} \sum_{m=1}^{M} \sum_{n=1}^{d} \big[ \vec{\Delta}_{m,n}^{T} \mathbf{\Sigma}_{m}^{-1} \vec{\Delta}_{m,n} + \ln (\vert \mathbf{\Sigma}_{m} \vert) \big]
\end{align}
with respect to the same one-dimensional parameters $a^{2}$ and $\sigma^{2}$ as before, due to the assumption of isotropic motion.  Here, the quantities $\smash{\vec{\Delta}_{m,n}}$ and $\smash{\mathbf{\Sigma}_{m}}$ are defined as before, except that they vary with the trajectory ($m=1,\ldots,M$) and dimension ($n=1,\ldots,d$), respectively.  In general, the minimization has to be conducted numerically, but thanks to the linear dependence of $\mathbf{\Sigma}$ on the parameters it becomes analytically tractable on the boundaries, where either $a^{2}$ or $\sigma^{2}$ becomes zero.  The remaining parameter can then be estimated as follows,
\begin{align}
\label{eq:mle-boundary-a2-solution}
a^{2} \vert_{\sigma^{2} = 0} & = \frac{1}{d \NM} \sum_{m=1}^{M} \sum_{n=1}^{d} \vec{\Delta}_{m,n}^{T} \mathbf{\Sigma}_{m}'^{-1} \vec{\Delta}_{m,n} \, ,
\\
\label{eq:mle-boundary-sigma2-solution}
\sigma^{2} \vert_{a^{2} = 0} & = \frac{1}{d \NM} \sum_{m=1}^{M} \sum_{n=1}^{d} \vec{\Delta}_{m,n}^{T} \mathbf{\Sigma}_{m}''^{-1} \vec{\Delta}_{m,n} \, ,
\end{align}
where $\smash{\NM = \sum_{m=1}^{M} N_{m}}$ denotes the total number of considered increments and $N_{m}$ is the number of entries in the vector $\vec{\Delta}_{m,n}$, independent of $n$.  Note that the vector-matrix products are all strictly positive, because the matrices $\mathbf{\Sigma'}$, $\mathbf{\Sigma''}$ and $\mathbf{\Sigma}$, as well as their inverses, are all positive definite (see further Appendix~\ref{app:positive-definiteness}).  If both $a^{2}$ and $\sigma^{2}$ are greater than zero, Eq.~\eqref{eq:M-log-likelihood} can effectively be reduced to a one-dimensional optimization problem in the vein of Ref.~\onlinecite{VestergaardBlainey2014}, where the sine-transformed counterpart of Eq.~\eqref{eq:log-likelihood} was considered.  We thereby introduce the new parameters $a^{2}$ and $\phi = \sigma^{2} / a^{2}$, which make the covariance matrix and the log-likelihood function separable, \emph{i.e.},~$\mathbf{\Sigma}_{m} = \smash{a^{2} \mathbf{\tilde{\Sigma}}_{m}(\phi)}$ $\forall m$ and
\begin{align}\label{eq:new-log-likelihood}
\notag
\mathcal{L}(\{ \vec{\Delta}_{m,n} \} \mid a^{2}, \phi) & = \frac{1}{2} \sum_{m=1}^{M} \sum_{n=1}^{d} \bigg[ \frac{1}{a^{2}} \vec{\Delta}_{m,n}^{T} \mathbf{\tilde{\Sigma}}_{m}^{-1} \vec{\Delta}_{m,n}
\\
& \mathrel{\phantom{=}} + N_{m} \ln (a^{2}) + \ln (\vert \mathbf{\tilde{\Sigma}}_{m} \vert) \bigg] \, .  
\end{align}
The determinants $\ln (\vert \mathbf{\tilde{\Sigma}}_{m} \vert)$ can be evaluated via Eqs.~\eqref{eq:general-determinant} and~\eqref{eq:q=0-determinant} by replacing $\alpha$ and $\beta$ with $\smash{\tilde{\alpha}} = 1 + \phi (1 - 2 B_{m})$ and $\smash{\tilde{\beta}} = -1/2 + \phi B_{m}$, respectively.  Equation~\eqref{eq:new-log-likelihood} gets minimized with respect to $a^{2}$ for
\begin{equation}\label{eq:general-a2-solution}
a^{2} = \frac{1}{d \NM} \sum_{m=1}^{M} \sum_{n=1}^{d} \vec{\Delta}_{m,n}^{T} \mathbf{\tilde{\Sigma}}_{m}^{-1} \vec{\Delta}_{m,n} \, ,
\end{equation}
and therefore reduces to
\begin{align}\label{eq:1D-log-likelihood}
\mathcal{L}(\{ \vec{\Delta}_{m,n} \} \mid \phi)
\notag
& = \frac{d \NM}{2} \ln \Bigg( \sum_{m=1}^{M} \sum_{n=1}^{d} \vec{\Delta}_{m,n}^{T} \mathbf{\tilde{\Sigma}}_{m}^{-1} \vec{\Delta}_{m,n} \Bigg)
\\
& \mathrel{\phantom{=}} + \frac{d}{2} \sum_{m=1}^{M} \ln (\vert \mathbf{\tilde{\Sigma}}_{m} \vert) \, ,
\end{align}
except for an additive constant $d \NM / 2 - d \NM \ln (d \NM) / 2$, which can be neglected because it is independent of $a^{2}$ and $\sigma^{2}$.  Minimizing Eq.~\eqref{eq:1D-log-likelihood} with respect to $\phi$ is a nonlinear one-dimensional optimization problem, which can be tackled conveniently using robust derivative-free algorithms, such as Brent's method~\cite{Brent1973} or golden-section search~\cite{Kiefer1953}.  The Julia data analysis package~\cite{JuliaScript} makes use of an implementation of Brent's method provided by the \texttt{Optim} package~\cite{MogensenRiseth2018}.  

\begin{figure}[t!]
\centering
\includegraphics{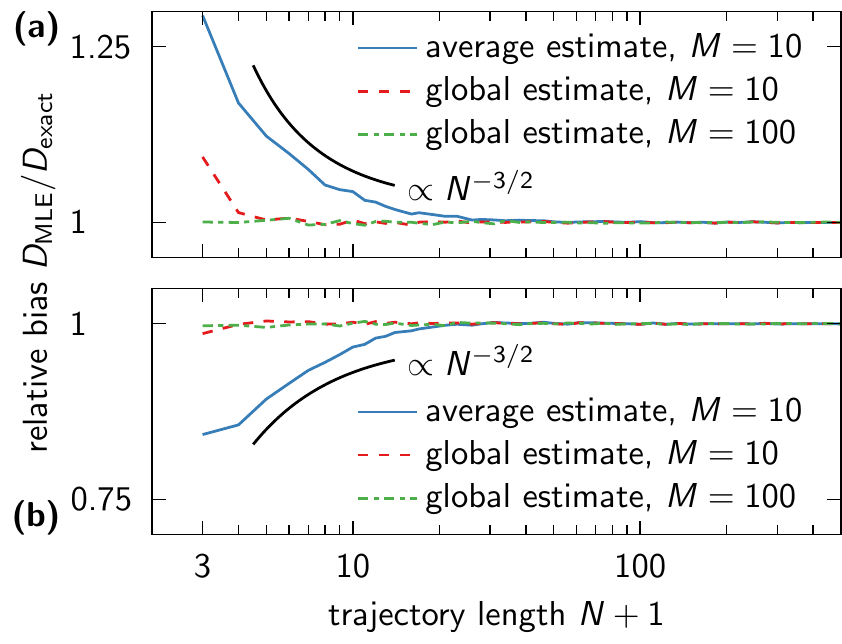}
\caption{Estimating relative bias in the diffusion coefficient MLE with respect to the trajectory length.  We analyzed $M$ simulated trajectories of equal lengths $N+1$, either separately or collectively, using Eqs.~\eqref{eq:mle-boundary-a2-solution}, \eqref{eq:mle-boundary-sigma2-solution}, \eqref{eq:general-a2-solution} and~\eqref{eq:1D-log-likelihood}.  We considered two cases: a signal-to-noise ratio of (a)~$\sigma^{2} / a^{2} = 1/2$ and (b)~$\sigma^{2} / a^{2} = 2$.   The resulting average diffusion coefficient (blue solid line), which was determined from the single-trajectory estimates and compared to the simulation input value $D_{\text{exact}}$, was either over or underestimated, respectively, for $N \lesssim 20$.  As indicated by the guides to the eye (solid black lines), the bias is $\mathcal{O} ( N^{-3/2} )$, which becomes $\mathcal{O} ( \smash{\NM^{-3/2}} )$ for the global MLE estimates (dashed red line for $M=10$, dashed-dotted green line for $M=100$).  This is why the latter are virtually unbiased for almost all trajectory lengths.  All results were averaged over multiple realizations to reduce noise.  }
\label{fig:mle_bias}
\end{figure}

To assess the uncertainty of the estimates, we turn to the standard errors $\smash{\delta \theta = \sqrt{\var(\theta)}}$ of the model parameters $\smash{\theta \in \{ a^{2}, \sigma^{2} \}}$.  These are bounded from below by the Cram\'{e}r-Rao bounds, which are computed from the Fisher information corresponding to the likelihood as
\begin{align}
\begin{split}\label{eqs:errors}
\delta a^{2} \vert_{\sigma^{2} = 0} & \geq \sqrt{\frac{2}{d \NM}} a^{2} \vert_{\sigma^{2} = 0} \, ,
\\
\delta \sigma^{2} \vert_{a^{2} = 0} & \geq \sqrt{\frac{2}{d \NM}} \sigma^{2} \vert_{a^{2} = 0} \, , 
\\
\delta a^{2} & \geq \sqrt{ \frac{I_{2,2}}{I_{1,1} I_{2,2} - I_{1,2}^{2}} } \, , 
\\
\delta \sigma^{2} & \geq \sqrt{ \frac{(a^{2})^{2} I_{1,1} - 2 \phi a^{2} I_{1,2} + \phi^{2} I_{2,2}}{I_{1,1} I_{2,2} - I_{1,2}^{2}} } \, .  
\end{split}
\end{align}
The matrix elements $I_{i,j}$ of the Fisher information matrix $\mathbf{I}(a^{2},\phi)$ are specified in Appendix~\ref{app:fisher-information} along with a detailed derivation of the above equations.  The bounds in Eqs.~\eqref{eqs:errors} become tight in the limit $\NM \to \infty$, because the estimators are asymptotically unbiased (see Fig.~\ref{fig:mle_bias}).  For extremely sparse datasets, where the lower bounds of Eqs.~\eqref{eqs:errors} vastly underestimate the standard errors, other methods have to be employed to estimate the uncertainty, such as bootstrapping.  

It is worth mentioning that the formalism of Refs.~\onlinecite{Berglund2010} and~\onlinecite{MichaletBerglund2012}, which treats the problem in discrete sine space (albeit only for $M=d=1$, but the extension to $d$ and $M$ different from one is straightforward), results in identical numerical values for the parameter estimates and standard errors.  However, our approach has the advantage that we obtain closed-form expressions for the standard errors [Eqs.~\eqref{eqs:errors}], whereas these same quantities involve finite sums that grow with the trajectory length in discrete sine space.  Working in real space can therefore be beneficial under certain circumstances, \emph{e.g.}, in the context of Bayesian inference involving priors based on the Fisher information matrix (see further Sec.~\ref{sec:bayesian-inference}).  

Finally, note that the estimators $a^{2}$ and $\phi$ (or, equivalently, $a^{2}$ and $\sigma^{2}$) are slightly correlated, as seen by the non-vanishing off-diagonal elements $I_{1,2} = I_{2,1}$ of the Fisher information matrix.  Figure~\ref{fig:mle_bias} explores the bias in estimates of $\sigma^{2}$ and the associated diffusion coefficient as a function of the trajectory length $N+1$, using two-dimensional trajectories generated via Brownian dynamics simulations.  The simulations were conducted using discretized versions of Eqs.~\eqref{eq:Y-process}--\eqref{eq:X-process} (see Appendix~\ref{app:brownian-dynamics-simulations}), and analyzed both on a single-trajectory level and collectively.  Our numerical results demonstrate that a bias of $\mathcal{O} ( N^{-3/2} )$ affects single-trajectory estimates of short trajectories, but is virtually non-existent in global estimates, where the bias scales like $\mathcal{O} ( \smash{\NM^{-3/2}} )$, \emph{i.e.}, with the total number of increments.  This is in good agreement with analytical results obtained in Ref.~\onlinecite{VestergaardBlainey2014}, which show that the estimators in discrete sine space are unbiased up to $\mathcal{O} ( N^{-1} )$.  However, Ref.~\onlinecite{VestergaardBlainey2014} claims that next to said bias, which originates from the asymmetry of the likelihood function, there should be an additional $\mathcal{O} ( N^{-1/2} )$ bias coming from the fact that the maximum likelihood approach requires $a^{2}$ and $\sigma^{2}$ to be positive.  The diffusion coefficient estimator in real space does not exhibit this second bias, at least not for the signal-to-noise values considered here.

\subsection{Quality factor analysis}\label{sec:quality-factor}

How can we test that we are actually observing free diffusion?  According to Eq.~\eqref{eq:pdf}, for a diffusive process the elements of $\smash{\mathbf{\Sigma}_{m}^{-1/2} \vec{\Delta}_{m,n}}$ for a fixed $n$ are independent, uncorrelated normal random variables with zero mean and unit variance.  The sum of their squares, given by
\begin{equation}\label{eq:quadratic-form}
\chi_{m}^{2} = \sum_{n=1}^{d} \vec{\Delta}_{m,n}^{T} \mathbf{\Sigma}_{m}^{-1} \vec{\Delta}_{m,n} > 0 \, , 
\end{equation}
should therefore be distributed according to a $\chi^{2}$-distribution with $d N_{m}$ degrees of freedom.  Note that we do not correct $d N_{m}$ by the number of model parameters, because the degrees of freedom associated with a single trajectory are just a tiny fraction of the overall number of unconstrained degrees of freedom, $d \NM - 2$.  

In principle, we could test whether a sample of $M$ trajectories adheres to the diffusion model in Eqs.~\eqref{eq:Y-process}--\eqref{eq:X-process} by verifying the distribution of the corresponding $\chi_{m}^{2}$-values, but only if they are all of equal length.  For trajectories that differ in their lengths $N_{m}+1$, the quadratic forms $\chi_{m}^{2}$ follow different $\chi^{2}$-distributions.  To simplify the analysis, we focus on their corresponding cumulative distribution functions, which are all uniform on the interval $[0,1)$.  The associated statistic is given by
\begin{equation}\label{eq:quality-factor}
\Qoppa_{m} = 1 - \frac{\gamma (d N_{m} / 2, \chi_{m}^{2} / 2)}{\Gamma (d N_{m} / 2)} \, ,
\end{equation}
where $\gamma(a,z)$ and $\Gamma(a)$ denote the lower incomplete and ordinary $\Gamma$-functions, respectively, that are defined as follows,
\begin{align*}
\gamma(a,z) = \int_{0}^{z} \ddf{x} x^{a-1} \e^{-x} \, , & & \Gamma(a) = \lim_{z \to \infty} \gamma(a,z) \, .  
\end{align*}
This is reminiscent of the discussion in Ref.~\onlinecite{Bullerjahnvon-Bulow2020}, where a quantity similar to the one in Eq.~\eqref{eq:quality-factor} is referred to as the quality factor $Q$, because its values are uniformly distributed whenever the elements of $\smash{\mathbf{\Sigma}_{m}^{-1/2} \vec{\Delta}_{m,n}}$ are truly independent.  The main difference between Eq.~\eqref{eq:quality-factor} and traditional quality factors, such as the one in Ref.~\onlinecite{Bullerjahnvon-Bulow2020}, is that the latter consider $\chi^{2}$-statistics computed by summing over the weighted squared differences between data and model predictions.  Here, the model makes predictions about the distribution of $\chi_{m}^{2}$ of individual trajectories $m$ and, in turn, of $\Qoppa_{m}$, into which the parameters $a^{2}$ and $\sigma^{2}$ enter via the covariance matrix.  To highlight this distinction, we make use of the archaic greek letter \emph{qoppa}, instead of $Q$, to denote this (somewhat unorthodox) quality factor.  

To test whether a set $\{ \Qoppa_{m} \}_{m=1, \dots, M}$ of quality factor values follows the uniform distribution on $[0,1)$ expected for diffusive processes, we employ a variation of the Kolmogorov-Smirnov statistic called the Kuiper statistic~\cite{Kuiper1960, Tygert2010}.  Kuiper's variant measures the largest vertical deviations of the empirical distribution function above and below the cumulative distribution function, and is defined as their sum.  In our case, it thus reads
\begin{equation}\label{eq:kuiper-statistic}
\kappa = \sqrt{M} \bigg( \max_{m} \bigg[\frac{m}{M} - \Qoppa_{m} \bigg] + \max_{m} \bigg[ \Qoppa_{m} - \frac{m-1}{M} \bigg] \bigg)
\end{equation}
for a list of $\Qoppa_{m}$-values, $m=1,2,\dots,M$, sorted in ascending order.  Ideally, we have $\kappa \approx 1$, but significantly larger $\kappa$-values arise when the underlying assumption of $\smash{\mathbf{\Sigma}_{m}^{-1/2} \vec{\Delta}_{m,n}}$ being normal distributed is violated.  This is, \emph{e.g.}, the case for heterogeneous data arising from subpopulations with differing diffusion coefficients, as discussed in Sec.~\ref{sec:em-algorithm}.  

Finally, it should be mentioned that one can construct a right-tailed \emph{p}-value from the sampling distribution of $\kappa$.  Asymptotically, it takes the form
\begin{equation}\label{eq:p-value}
p \mathop{\sim}^{M \to \infty} 2 \sum_{m=1}^{\infty} (4 m^{2} \kappa^{2} - 1) \e^{-2 m^{2} \kappa^{2}} \, ,
\end{equation}
where the sum has to be truncated at some large value, \emph{e.g.}, at $m=100$ or $m=1000$, if used in practice.  The associated probability $p$ can be helpful to determine whether a $\kappa$-value is reasonable or not.  For sparse datasets, tables can be used to look up the \emph{p}-value~\cite{Arsham1988}.

\subsection{Analysis of mixtures of subpopulations}\label{sec:em-algorithm}

In situations where a single diffusion coefficient cannot describe the observed dynamics in a sample of trajectories, the single-population analysis of Sec.~\ref{sec:mle} breaks down.  Here, we extend the model by introducing $K$ distinct subpopulations, each characterized by a set of parameters $\smash{\{ a^{2}_{k}, \sigma^{2}_{k} \}}$, $k = 1,2, \dots, K$.  Every tracked particle is assumed to belong to one of these subpopulations, but the exact assignment is not known \emph{a priori}.  Particles are not allowed to switch between populations and therefore have a fixed diffusion coefficient, depending on which subpopulation $k$ they belong to.  The task of parameter fitting now becomes two-fold: The parameters of each subpopulation have to be varied to find their optimal values, while, at the same time, the particle trajectories have to be assigned to the subpopulations they most likely belong to.  Problems like these can be treated with the help of the EM algorithm~\cite{DempsterLaird1977}, as outlined below.  

To analyze data from a mixture of $K$ subpopulations with distinct diffusive dynamics, we consider the following likelihood function with latent mixing fractions $P_{k}$,
\begin{align}
\notag
\ell(\{ \vec{\Delta}_{m,n} \} \mid \{ & P_{k}, a_{k}^{2}, \sigma_{k}^{2} \}_{k=1,\dots,K})
\\
& = \prod_{m=1}^{M} \sum_{k=1}^{K} P_{k} \prod_{n=1}^{d} \ell_{k}(\vec{\Delta}_{m,n} \mid a_{k}^{2}, \sigma_{k}^{2}) \, .  
\end{align}
The one-dimensional likelihoods $\smash{\ell_{k}(\vec{\Delta}_{m,n} \mid a_{k}^{2}, \sigma_{k}^{2})}$ are given by Eq.~\eqref{eq:pdf}, where the subscript $k$ indicates that the covariance matrix $\mathbf{\Sigma} = \mathbf{\Sigma}_{m}$ is evaluated using the subpopulation parameters $\smash{a_{k}^{2}}$ and $\smash{\sigma_{k}^{2}}$.  Now, due to the sum appearing in $\smash{\ell(\{ \vec{\Delta}_{m,n} \} \mid \{ P_{k}, a_{k}^{2}, \sigma_{k}^{2} \})}$, its negative log-likelihood remains intractable.  This is where the EM algorithm comes in: Instead of minimizing $- \ln (\ell)$ explicitly, we consider the upper bound
\begin{align}\label{eq:upper-bound}
\notag
L (\{ & \vec{\Delta}_{m,n} \} \mid \{ P_{k}, a_{k}^{2}, \sigma_{k}^{2} \})
\\
\notag
& = - \sum_{m=1}^{M} \sum_{k=1}^{K} T_{k,m} \ln \Bigg( \frac{P_{k}}{T_{k,m}} \prod_{n=1}^{d} \ell_{k}(\vec{\Delta}_{m,n} \mid a_{k}^{2}, \sigma_{k}^{2}) \Bigg)
\\
\notag
& \geq - \sum_{m=1}^{M} \ln \Bigg( \sum_{k=1}^{K} T_{k,m} \frac{P_{k}}{T_{k,m}} \prod_{n=1}^{d} \ell_{k}(\vec{\Delta}_{m,n} \mid a_{k}^{2}, \sigma_{k}^{2}) \Bigg)
\\
& = - \ln \big( \ell(\{ \vec{\Delta}_{m,n} \} \mid \{ P_{k}, a_{k}^{2}, \sigma_{k}^{2} \}) \big) \, ,
\end{align}
which follows from Jensen's inequality.  In principle, such upper bounds can be constructed for arbitrary coefficients $T_{k,m} > 0$, whose interpretation becomes clearer in what follows.  If we were able to choose the $T_{k,m}$ such that equality holds in Eq.~\eqref{eq:upper-bound} for all parameter tuples $\{ P_{k}, \smash{a^{2}_{k}, \sigma^{2}_{k}} \}_{k=1, \dots, K}$, then it would not matter if we minimized $- \ln (\ell)$ or $L$, given a known set of  $T_{k,m}$.  This is exactly the case whenever
\begin{equation*}
\frac{P_{k}}{T_{k,m}} \prod_{n=1}^{d} \ell_{k}(\vec{\Delta}_{m,n} \mid a_{k}^{2}, \sigma_{k}^{2}) = \const \, ,
\end{equation*}
so by requiring that the $T_{k,m}$ are also normalized, we finally obtain
\begin{align}\label{eq:T-coefficients}
T_{k,m}
\notag
& = \Bigg[ \sum_{k'=1}^{K} P_{k'} \prod_{n=1}^{d} \ell_{k'}(\vec{\Delta}_{m,n} \mid a_{k}^{2}, \sigma_{k}^{2}) \Bigg]^{-1}
\\
& \mathrel{\phantom{=}} \times P_{k} \prod_{n=1}^{d} \ell_{k}(\vec{\Delta}_{m,n} \mid a_{k}^{2}, \sigma_{k}^{2}) \, .  
\end{align}
It now becomes apparent that $T_{k,m}$ corresponds to the probability of trajectory $m$ belonging to the $k$-th subpopulation.  

The EM algorithm is a two-step algorithm, where in the first step the current values of $\{ P_{k}, a_{k}^{2}, \sigma_{k}^{2} \}$ are used to update the classification coefficients $T_{k,m}$ via Eq.~\eqref{eq:T-coefficients}.  In the second step, the latter are plugged into the upper bound [Eq.~\eqref{eq:upper-bound}], which can be rewritten as
\begin{align}\label{eq:em-log-likelihood}
\notag
L ( \{ \vec{\Delta}_{m,n} & \} \mid \{ P_{k}, a_{k}^{2}, \sigma_{k}^{2} \} ) = \sum_{m=1}^{M} \sum_{k=1}^{K} T_{k,m} 
\\
& \times \Bigg[ \sum_{n=1}^{d} \mathcal{L}(\vec{\Delta}_{m,n} \mid a_{k}^{2}, \sigma_{k}^{2}) - \ln \bigg( \frac{P_{k}}{T_{k,m}} \bigg) \Bigg] \, .  
\end{align}
Here, $\smash{\mathcal{L}(\vec{\Delta}_{m,n} \mid a_{k}^{2}, \sigma_{k}^{2})}$ is defined in Eq.~\eqref{eq:log-likelihood} for $\mathbf{\Sigma} = \mathbf{\Sigma}_{m}$.  Minimizing Eq.~\eqref{eq:em-log-likelihood} with respect to $\{ P_{k}, a_{k}^{2}, \sigma_{k}^{2} \}$ updates the estimate for the parameters, and the first step is repeated.  Said minimization can mostly be done analytically, because Eq.~\eqref{eq:em-log-likelihood} is separable, and therefore susceptible to the methods applied in Sec.~\ref{sec:mle}.  For the mixing fractions, we get
\begin{equation}\label{eq:mixing-fractions}
P_{k} = \frac{1}{M} \sum_{m=1}^{M} T_{k,m} \, .  
\end{equation}
On the boundary, the solutions read
\begin{align}
\label{eq:em-boundary-a2-solution}
a_{k}^{2} \big\vert_{\sigma_{k}^{2} = 0} & = \frac{1}{C_{k}} \sum_{m=1}^{M} T_{k,m} \sum_{n=1}^{d} \vec{\Delta}_{m,n}^{T} \mathbf{\Sigma}_{m}'^{-1} \vec{\Delta}_{m,n} \, ,
\\
\label{eq:em-boundary-sigma2-solution}
\sigma_{k}^{2} \big\vert_{a_{k}^{2} = 0} & = \frac{1}{C_{k}} \sum_{m=1}^{M} T_{k,m} \sum_{n=1}^{d} \vec{\Delta}_{m,n}^{T} \mathbf{\Sigma}_{m}''^{-1} \vec{\Delta}_{m,n} \, ,
\\
\notag
C_{k} & = d \sum_{m=1}^{M} T_{k,m} N_{m} \, .  
\end{align}
Using the parameter representation $\smash{(a_{k}^{2}, \phi_{k})}$ with $\smash{\phi_{k} = \sigma_{k}^{2} / a_{k}^{2}}$, Eq.~\eqref{eq:em-log-likelihood} is minimized for
\begin{align}
\label{eq:em-general-a2-solution}
a_{k}^{2} & = \frac{1}{C_{k}} \sum_{m=1}^{M} T_{k,m} \sum_{n=1}^{d} \vec{\Delta}_{m,n}^{T} \mathbf{\tilde{\Sigma}}_{m}^{-1} \vec{\Delta}_{m,n} \, ,
\\
\notag
\phi_{k} & = \argmin_{\phi_{k}} \frac{C_{k}}{2} \ln \Bigg( \sum_{m=1}^{M} T_{k,m} \sum_{n=1}^{d} \vec{\Delta}_{m,n}^{T} \mathbf{\tilde{\Sigma}}_{m}^{-1} \vec{\Delta}_{m,n} \Bigg)
\\
\label{eq:em-general-sigma2-solution}
& \mathrel{\phantom{ = \argmin_{\phi_{k}}}} + \frac{d}{2} \sum_{m=1}^{M} T_{k,m} \ln (\vert \mathbf{\tilde{\Sigma}}_{m} \vert) \, .  
\end{align}
The high-dimensional minimization problem therefore reduces to a series of analytic expressions [Eqs.~\eqref{eq:mixing-fractions}, \eqref{eq:em-boundary-a2-solution}, \eqref{eq:em-boundary-sigma2-solution} and~\eqref{eq:em-general-a2-solution}], and a handful of independent one-dimensional optimization problems [Eq.~\eqref{eq:em-general-sigma2-solution}], which can be solved numerically in an efficient way.  The algorithm is implemented in a Julia data analysis package~\cite{JuliaScript}.  

Finally, we would like to mention that one could also consider a continuum of $a^{2}$ and $\sigma^{2}$ values, as an alternative to discrete subpopulations.  If only $a^{2}$ is allowed to vary between trajectories, a latent variable model could be considered.  In general, a Bayesian formulation would naturally lend itself to a continuous-parameter treatment, in which a prior distribution would be updated into a posterior distribution in light of the observed data.

\subsection{Selection criterion for optimal number of subpopulations}\label{sec:selection-criterion}

\begin{figure}[t!]
\centering
\includegraphics{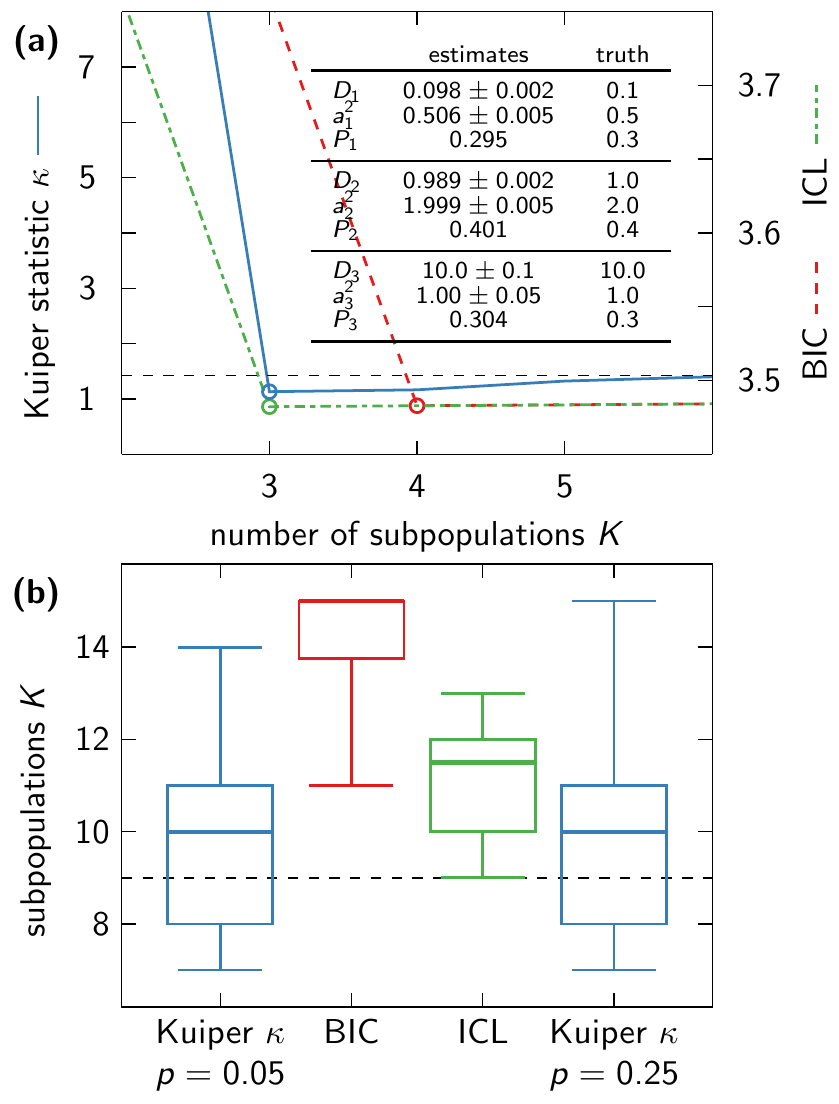}
\caption{Verifying the effectiveness of quality factor statistics as a selection criterion.  (a)~Determining the optimal number $K$ of subpopulations.  We generated $300 + 400 + 300 = 1000$ synthetic trajectories of different lengths (uniformly distributed on $[4,101]$) via Brownian dynamics simulations, using the parameter values listed in the table under ``truth''.  After analyzing the data with the EM algorithm for $K=1,2, \dots, 6$, the corresponding Kuiper statistic [Eq.~\eqref{eq:kuiper-statistic}, solid blue line] was computed.  For $K \geq 3$, $\kappa$ remained below $\kappa \approx 1.42$ (dashed black line), corresponding to $p$-values greater than $0.25$.  Accordingly, we chose $K=3$ (blue circle) as the optimal number of subpopulations.  The associated parameter estimates are listed in the table under ``estimates'', and are in excellent agreement with ground truth.  The red and green circles denote the corresponding BIC (dashed red line) and ICL (dashed-dotted green line) predictions.  (b)~Statistics of predictions for 20 distinct datasets, each containing $110 + 30 + 70 + 80 + 190 + 230 + 110 + 120 + 60 = 1000$ trajectories sampled from a complex mixture of nine subpopulations ($\smash{a_{k=1,2,\dots,9}^{2}} = 0.04$, $0.04$, $0.05$, $0.13$, $0.06$, $0.09$, $0.20$, $0.35$, $0.99$, and $\smash{\sigma_{k=1,2,\dots,9}^{2}} = 0.08$, $0.0005$, $0.009$, $0.02$, $0.18$, $0.32$, $0.10$, $0.34$, $0.72$) and following the same length distribution as in (a).  The data were analyzed for $K=1,2, \dots, 15$, and the predictions of BIC (red) and ICL (green) were determined by the positions of their minima on said interval.  In contrast, the novel selection criterion (blue) relied on the first instance where $\kappa$ went below a threshold value of either $\kappa \approx 1.75$ ($p = 0.05$) or $\kappa \approx 1.42$ ($p = 0.25$).  In the event of the threshold not being reached, which happened more frequently for the lower threshold, the position where $\kappa$ got minimized was chosen.  }
\label{fig:proof_of_principle}
\end{figure}

As the number of subpopulations $K$ is increased, the fit to heterogeneous data gradually gets better.  Because the regularity conditions for conventional criteria, such as the Bayesian information criterion (BIC), do not hold for finite mixture models~\cite{BiernackiCeleux2000}, we propose to rely on a quality factor analysis for model selection.  This is done as follows: After repeated fitting of a dataset via the EM algorithm for a fixed $K$, the classification coefficients $T_{k,m}$ that (along with the optimal parameter values $\smash{\{ P_{k}, a_{k}^{2}, \sigma_{k}^{2} \}}$) minimize Eq.~\eqref{eq:em-log-likelihood} are used to assign each trajectory $m$ to the subpopulation $k$ it most likely belongs to, according to
\begin{equation}
k = \argmax_{i} T_{i,m} \, .  
\end{equation}
The associated quality factors $\smash{\{ \Qoppa_{m}^{(k)} \}}$ are then determined using the subpopulation-specific parameters $\smash{a_{k}^{2}}$ and $\smash{\sigma_{k}^{2}}$, resulting in a set $\big\{ \smash{\{ \Qoppa_{m}^{(1)} \}}, \smash{\{ \Qoppa_{m}^{(2)} \}}, \dots, \smash{\{ \Qoppa_{m}^{(K)} \}} \big\}$ that is finally plugged into Eq.~\eqref{eq:kuiper-statistic} to evaluate the Kuiper statistic $\kappa$.  Figure~\ref{fig:proof_of_principle}(a) serves as a proof of concept for this selection procedure, where $\kappa$ is plotted as a function of $K$ for a simulated heterogeneous dataset with three distinct subpopulations.  The estimates for $K=3$ are in excellent agreement with the input parameter values of our simulations for three subpopulations, and result in $\kappa \approx 1.13$.  Note that the slight increase in $\kappa$ for $K > 3$ is due to stochasticity in the EM optimization, which becomes more pronounced as the number of subpopulations is increased (for details on our implementation of the EM algorithm, see Appendix~\ref{app:em-algorithm}).  Also note that we are not interested in the global minimum of $\kappa$ with respect to $K$, but in the smallest $K$ where $\kappa \approx 1$.  In practice, this can be realized by a threshold value for $\kappa$, below which $\kappa$ is considered sufficiently close to one to accept the associated $K$ as the optimal number of subpopulations.  We recommend practitioners to choose a threshold somewhere between $\kappa \approx 1.75$ and $1.42$ (corresponding to the $p$-values $0.05$ and $0.25$, respectively) that reflects their confidence in the quality of the data.  

For comparison, we show predictions of two established selection criteria, namely the BIC and the integrated completed likelihood (ICL) criterion~\cite{BiernackiCeleux2000}, in Fig.~\ref{fig:proof_of_principle}(a).  The former can be evaluated for our model as follows,
\begin{equation}
\operatorname{BIC} = \frac{2 L (\{ \vec{\Delta}_{m,n} \} \mid \{ P_{k}, a_{k}^{2}, \sigma_{k}^{2} \}) + (3 K - 1) \ln (d \mathcal{N}_{M})}{\mathcal{N}_{M}} \, ,
\end{equation}
where the likelihood is given by Eq.~\eqref{eq:em-log-likelihood}.  The ICL is formulated in an almost identical manner as the BIC, except that the classification coefficients $T_{k,m}$ in front of the square brackets in Eq.~\eqref{eq:em-log-likelihood} are replaced by
\begin{equation*}
\tilde{T}_{k,m} = \begin{cases}
1 & \argmax_{i} T_{i,m} = k \\
0 & \text{otherwise}
\end{cases} \, .  
\end{equation*}
Whereas the BIC overestimates the number of subpopulations, the ICL and the Kuiper statistics criterion produce reliable results for the mixture of three subpopulations. However, for more complex mixtures, also the ICL fails.  Figure~\ref{fig:proof_of_principle}(b) visualizes the prediction statistics of BIC, ICL and the Kuiper statistics criterion when applied to datasets composed of trajectories sampled from nine distinct subpopulations.  Although the distributions of predicted $K$-values are fairly broad, probably due to the trajectories being so short and few, the biases of BIC and ICL are clear, thus confirming the claim that traditional selection criteria cannot be relied on in our particular case.

\section{Application to empirical data}

\begin{figure*}[t!]
\centering
\includegraphics{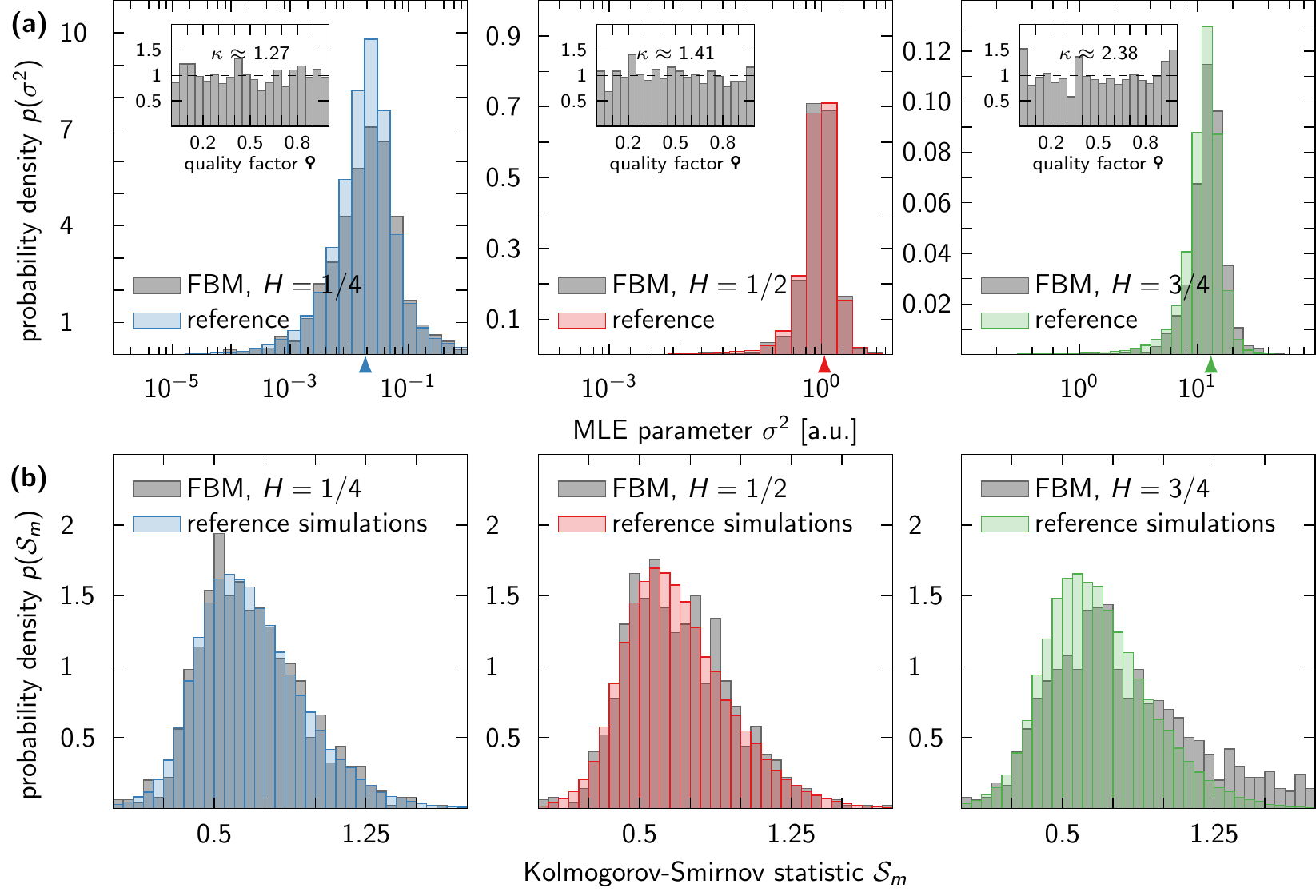}
\caption{Testing for diffusivity in FBM subject to static and dynamic noise.  Using $B=1/6$ and $a^{2} = \sigma^{2} = 1$, where $\sigma^{2}$ is the variance of the normal random variable used as input for the circulant method, we generated 1000 trajectories with lengths uniformly distributed on [4,101] governed by subdiffusive (left panels), diffusive (center panels) and superdiffusive (right panels) dynamics, and compared them to reference data made up of $10^5$ diffusive trajectories.  We considered the (a)~distributions of single-trajectory $\sigma^{2}$-estimates, distributions of quality factors calculated for the FBM data (insets), and (b)~distributions of KS statistics, which test for varying diffusivity within the trajectories.  The gray histograms are associated with the FBM data, whereas the colored histograms belong to the reference data.  The arrows at the bottom of the plots in (a) indicate the global $\sigma^{2}$-estimates.  For the superdiffusive data, we saw clear deviations in all three distributions, thus confirming that the process under scrutiny was non-diffusive.  However, for $H=1/4$ the results were more subtle: we only detected clearly noticeable deviations in the distribution of $\sigma^{2}$, which, in combination with the fact that the other two distributions fit the reference data, hinted at non-diffusive behavior.  }
\label{fig:fractional_brownian_motion}
\end{figure*}

\begin{figure*}[t!]
\centering
\includegraphics{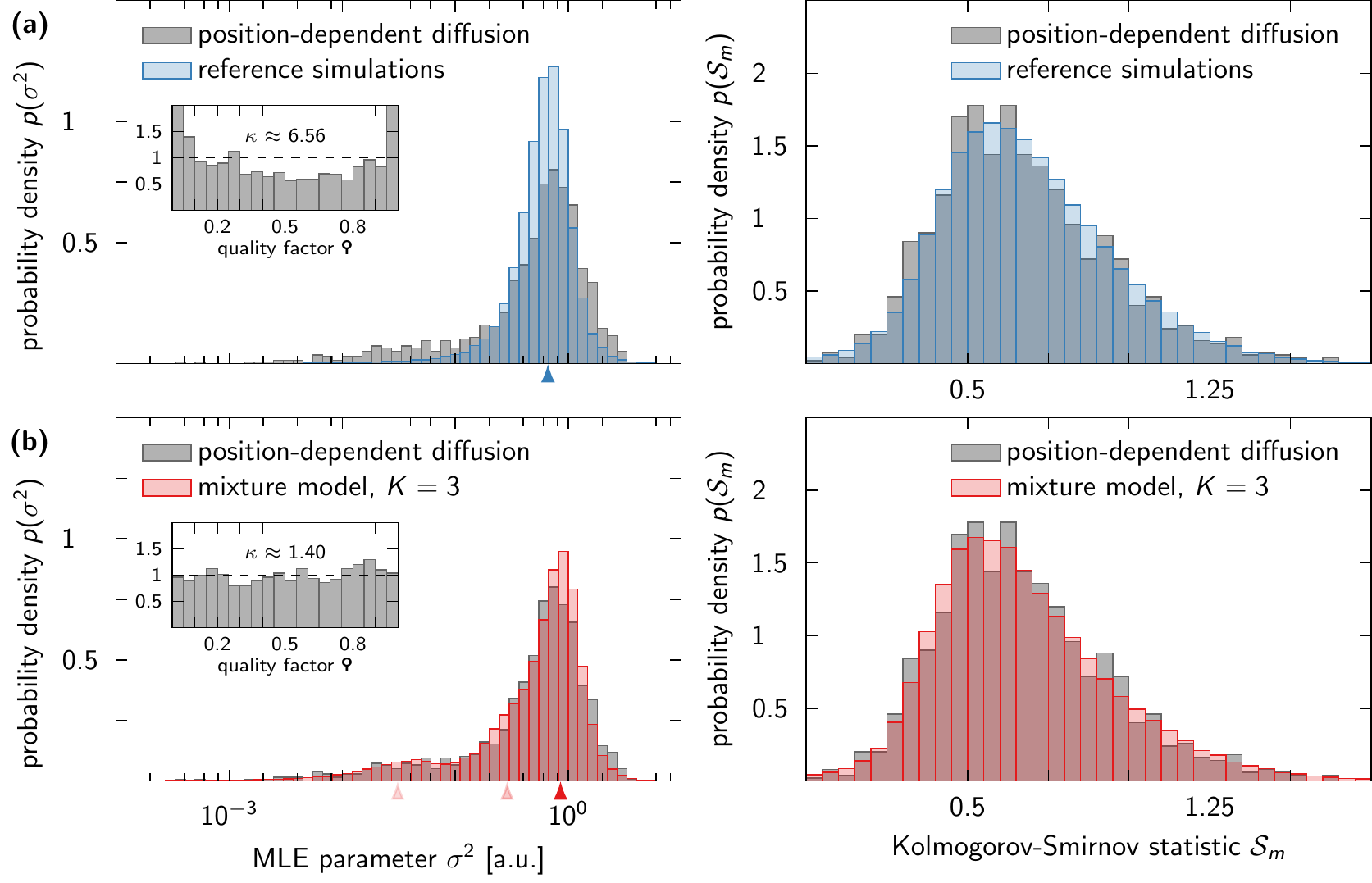}
\caption{Comparing position-dependent diffusion to mixture models.  (a)~Analogous to Fig.~\ref{fig:fractional_brownian_motion}, we compared PDD data (gray histograms) to diffusive reference data (blue histograms) in terms of their distributions of single-trajectory $\sigma^{2}$-estimates (left panel), quality factors (inset) and KS statistics (right panel).  The PDD data was generated via a stochastic process $X$ satisfying Eqs.~\eqref{eq:spatially-varying-diffusion}, \eqref{eq:Z-process} and~\eqref{eq:X-process}, resulting in 1000 non-diffusive trajectories with lengths uniformly distributed on [4,101].  The reference data consisted of $10^5$ trajectories following the same length distribution as the non-diffusive data, and were generated using $B=1/6$, and $a^{2}$ and $\sigma^{2}$ extracted from a global analysis of the PDD data (arrow indicates global $\sigma^{2}$-estimate).  The non-uniform quality factor distribution and strong deviations between the $\sigma^{2}$-distributions clearly indicate a non-diffusive behavior in the PDD data.  (b)~After analyzing the non-diffusive data via mixture models, we used the parameters for the minimal model best describing the data in (a) (arrows indicate the $\sigma^{2}$-estimates, and their relative opacities the respective mixing fractions) to generated the $10^5$ trajectories behind the red histograms.  \emph{Insets:}~Quality factor distributions for the PDD data.  The associated $\Qoppa$-values [Eq.~\eqref{eq:quality-factor}] were calculated using (a)~the parameters of a single-population fit, and (b)~the fit parameters of a mixture model with $K=3$.  }
\label{fig:inhomogeneous_diffusion}
\end{figure*}

To illustrate the use and effectiveness of the above theory, we applied it to simulations and experimental data.  Comparisons to the former are primarily intended to gauge the limits of the theoretical framework, while the analysis of experimental data is meant to demonstrate its use in practice.

\subsection{Simulated diffusion data}\label{sec:simulation-data}

As mentioned earlier, there are multiple modes of transport possible in microbiological systems, and a diffusion coefficient estimate is only reliable when the data satisfy the assumptions of the model.  In this regard, we generated one-dimensional synthetic trajectories imitating fractional Brownian motion (FBM)~\cite{Mandelbrotvan-Ness1968} and diffusion in inhomogeneous media~\cite{van-Kampen1988}, two processes which deviate from regular diffusion distinctly, and tested whether the theory was able to detect their non-diffusive behavior.  This was realized by analyzing three well-defined quantities: the distribution of quality factors, the distribution of single-trajectory $\sigma^{2}$-estimates, and the distribution of statistics resulting from a two-sample Kolmogorov-Smirnov (KS) test, which estimates how likely it is for early parts of a trajectory to be governed by the same diffusion dynamics as its later parts.  Said distributions were extracted from the non-diffusive data and compared visually to their diffusive counterparts.  

FBM is a Gaussian process with stationary increments, similar to the Wiener process in Eq.~\eqref{eq:Y-process}.  However, the increments of FBM need not be independent, and therefore give rise to a MSD that grows proportional to $\smash{t^{2H}}$ with $H \in (0,1)$ being the so-called Hurst index.  For $H=1/2$ we retrieve the Wiener process, whereas the cases $H < 1/2$ and $H > 1/2$ result in sub- and superdiffusion, respectively, due to negative and positive correlations.  We generated realizations of FBM via the circulant method~\cite{WoodChan1994}, and added static and dynamic noise by replacing $Y(t)$ in Eqs.~\eqref{eq:Y-process}--\eqref{eq:X-process} (or their discretized counterparts, see Appendix~\ref{app:brownian-dynamics-simulations}) with said realizations.  The resulting trajectories were analyzed both individually and globally using Eqs.~\eqref{eq:mle-boundary-a2-solution}, \eqref{eq:mle-boundary-sigma2-solution}, \eqref{eq:general-a2-solution} and~\eqref{eq:1D-log-likelihood}, where the solution that resulted in the lowest negative log-likelihood value was chosen in each case.  We used the global estimates for $a^{2}$ and $\sigma^{2}$, on the one hand, to calculate the corresponding $\Qoppa_{m}$-values [Eq.~\eqref{eq:quality-factor}] and, on the other hand, as input parameters for simulations of regular diffusion, \emph{i.e.} Eqs.~\eqref{eq:Y-process}--\eqref{eq:X-process}, which we considered as reference data.  For further details on the Brownian dynamics simulations, see Appendix~\ref{app:brownian-dynamics-simulations}.  The above-mentioned KS test was achieved by splitting $\smash{\mathbf{\Sigma}_{m}^{-1/2} \vec{\Delta}_{m}}$ in half (dropping the last entry if the number of elements in $\smash{\vec{\Delta}_{m}}$ was uneven), and forming two empirical distribution functions that were then compared.  Note that $d=1$, so we have dropped the index $n$ in $\smash{\vec{\Delta}_{m,n}}$.  The covariance matrices $\mathbf{\Sigma}_{m}$ were evaluated using the global parameter estimates.  The above procedures gave rise to the samples $\{ \sigma_{m}^{2} \}$, $\{ \Qoppa_{m} \}$ and $\{ \mathcal{S}_{m} \}$ for $m = 1,2, \dots, M$ of single-trajectory $\sigma^{2}$-estimates, quality factors and KS statistics, respectively, whose distributions are plotted in Fig.~\ref{fig:fractional_brownian_motion}.  Here and in the following, reference simulations were sampled extensively such that the statistical scatter in the reference histograms is negligible.  

Overall, the data sparsity and strong static noise, which in most cases lead to a low effective signal-to-noise ratio $\sigma^{2} / a^{2}$, made the non-diffusive nature of FBM somewhat difficult to detect.  For example, in the case of $H = 1/4$, the only significant discrepancy between the data and reference was found in the distribution of single-trajectory estimates of $\sigma^{2}$.  Yet, this turned out to be a clear indication for non-diffusive dynamics in light of the fact that the other two quantities did not deviate from their diffusive counterparts.  For $H = 3/4$, where discrepancies were found in all considered quantities, the situation was more definitive.  A possible reason for the vastly different outcomes for $H=1/4$ and $H=3/4$ is that the latter case results in a slower decay of correlations between increments than $H=1/4$.  The two cases are therefore by no means symmetric, as one might naively assume.  

The second non-diffusive process we considered allows for a position-dependent diffusion (PDD) profile $D(z) = \sigma(z)^{2} / 2 \Delta t$.  One thereby has to replace Eq.~\eqref{eq:Y-process} with the following stochastic differential equation,
\begin{equation}\label{eq:spatially-varying-diffusion}
\dot{Y}(t) = \frac{\sigma(Y) \sigma'(Y)}{\Delta t} + \frac{\sigma(Y)}{\sqrt{\Delta t}} \xi(t) \, ,
\end{equation}
which is to be interpreted in the sense of It\^{o}.  Here, $f'(z)$ denotes the derivative of $f(z)$ with respect to $z$.  The resulting process is generally not Gaussian, due to the presence of multiplicative noise.  For our simulations, we chose $\sigma(z) = \sigma_{0} [ 1 - \smash{ 0.9 \e^{-(z / z_{0})^{2} / 2}} ]$ to mimic the diffusion around a diffusivity well, and the initial positions $Y_{0}$ were drawn from a uniform distribution on the interval $[-5 z_{0}, 5 z_{0}]$.  Analogous to the FBM data, the lengths of the trajectories were uniformly distributed on $[4,101]$ and a total of 1000 non-diffusive trajectories were considered.  

Our results for $a^{2} = \sigma_{0}^{2} = 1$ and $z_{0} = 3$ are given in Fig.~\ref{fig:inhomogeneous_diffusion}(a).  Again, we compared our results to reference data simulated using the global MLE estimates of $a^{2}$ and $\sigma^{2}$ obtained from an analysis of the PDD data.  The fact that the quality factor distribution is anything but uniform clearly indicates that the process cannot be described by a single diffusion coefficient.  This is also mirrored in the lack of overlap between the PDD data and the reference seen in the distribution of single-trajectory $\sigma^{2}$-estimates.  Only the KS statistics seemed consistent between the two datasets.  We also fitted the non-diffusive data with mixture models and found a decent match ($\kappa \approx 1.40$) for $K = 3$ and the parameters $\smash{a_{k=1,2,3}^{2}} \approx 0.97$, $1.03$, $1.03$, $\smash{\sigma_{k=1,2,3}^{2}} \approx 0.29$, $0.86$, $0.03$, and $\smash{P_{k=1,2,3}} \approx 0.17$, $0.71$, $0.12$.  As a test for consistency and to see if diffusion in inhomogeneous media can be distinguished from heterogeneous mixtures using samples of short trajectories, we then used these parameters to generate new synthetic data.  These data, drawn from the mixture model best describing the PDD, are compared to the PDD data in Fig.~\ref{fig:inhomogeneous_diffusion}(b).  Overall, the data from the mixture model and PDD process are fairly similar, thus making it impossible to distinguish between the processes, at least for the PDD profile considered here.  

In conclusion, it can be challenging to distinguish between diffusive and non-diffusive processes when the samples are made up of extremely short trajectories.  Also the presence of static and dynamic noise muddies the waters, especially when the (effective) signal-to-noise ratio is poor.  We recommend that practitioners investigate the three above-mentioned distributions, compare their results to Brownian dynamics simulations, and even consider further tests, such as the goodness-of-fit test proposed in Ref.~\onlinecite{VestergaardBlainey2014}.

\subsection{Experimental data}\label{sec:experimental-data}

We also applied the above theory to single-molecule tracking data for human MET receptor tyrosine kinase, bound to labeled 3H3-Fab antibody fragments within the plasma membrane of HeLa cells~\cite{HarwardtYoung2017}.  In the original study, tracking in two dimensions was conducted for multiple cells using the super-resolution imaging protocol uPAINT~\cite{GiannoneHosy2010} and imaging fluorescence correlation spectroscopy~\cite{BagWohland2014}.  Next to ``resting'' MET, realized with the above-mentioned Fab antibodies, the authors of Ref.~\onlinecite{HarwardtYoung2017} also considered internalin B-bound MET.  Here, we limit our analysis to the Fab-data recorded via uPAINT for 10 randomly chosen cells.  To account for the fact that each cell might be influenced by its local environment, we refrained from pooling together data measured in different cells.  The temporal resolution of the experiment was $\Delta t = \SI{0.02}{\second}$, and we assumed a blurring coefficient of $B = 1/6$ for all trajectories.  

\begin{figure}[t!]
\centering
\includegraphics{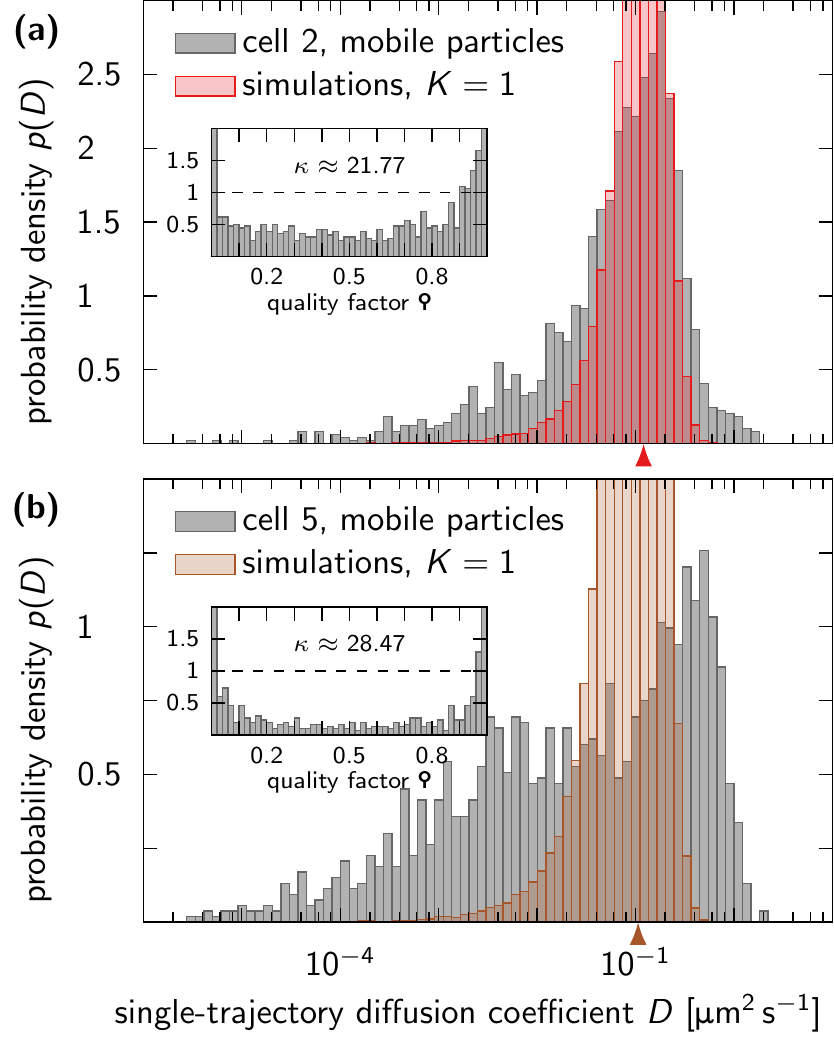}
\caption{Assessment of model with a single diffusion coefficient against single-molecule localization microscopy experiments.  Shown are the distributions of single-trajectory diffusion coefficients estimated from experiment (gray histograms) and from synthetic trajectories generated via a single-population model (colored histograms).  (a)~The global fit parameters $a^{2} = \SI{3.51(3)e-3}{\micro \meter \squared}$ and $D = \SI{1.21(1)e-1}{\micro \meter \squared \per \second}$ (red arrow) were used to simulate $10^{5}$ independent trajectories following the same length distribution as the experimental data for cell 2.  The distribution of single-trajectory diffusion coefficients (red histogram) turned out to be too narrow to explain the experimental observations.  (b)~Same as in (a), except that for cell 5 the parameters underlying the brown histogram are $a^{2} = \SI{3.09(3)e-3}{\micro \meter \squared}$ and $D = \SI{1.062(9)e-1}{\micro \meter \squared}$ (brown arrow).  \emph{Insets:} Distributions of quality factors corresponding to the experimentally observed trajectories, evaluated using the respective global single-population fit parameters.  Their horned shapes strongly deviate from the expected uniform distribution on $[0,1)$ (black dashed lines), which highlights the fact that the experimental data cannot be properly described by a single-population model.  }
\label{fig:single_population}
\end{figure}

Each trajectory of a given cell was first analyzed separately, analogous to the simulation data in Sec.~\ref{sec:simulation-data}.  This gave rise to $M$ diffusion coefficients per cell ($M = 1280$, $1943$, $1567$, $1709$, $1840$, $874$, $1165$, $1582$, $1178$, $1357$ for cells 1 through 10), \emph{i.e.}, one for each trajectory, and allowed us to sieve out the trajectories of essentially immobile particles, which we defined as those having $D < \smash{\SI{1e-6}{\micro \meter \squared \per \second}}$.  This is in line with the original analysis of the data~\cite{HarwardtYoung2017} but, alternatively, one can explicitly account for the immobile fraction in the modeling~\cite{SchutzSchindler1997}.  We then analyzed globally the remaining trajectories to estimate $\smash{a^{2}}$ and $\smash{\sigma^{2}}$ for each cell.  The normalized histograms of single-trajectory diffusion coefficients for two distinct cells are depicted in Fig.~\ref{fig:single_population}, next to Brownian dynamics simulation results for a single-population model.  The simulations were conducted using discretized versions of Eqs.~\eqref{eq:Y-process}--\eqref{eq:X-process} (see Appendix~\ref{app:brownian-dynamics-simulations}), where $\smash{a^{2}}$ and $\smash{\sigma^{2}}$ were set equal to the global cell parameter estimates, and the length of each trajectory was drawn from the observed length distribution of the respective cell.  

Figure~\ref{fig:single_population} clearly illustrates that the experimentally determined distributions are too broad to be explained by a single set of diffusion parameters.  To verify this conclusion, we computed the corresponding quality factors [Eq.~\eqref{eq:quality-factor}], one for each considered trajectory, and inspected visually, as well as with the help of the Kuiper statistic [Eq.~\eqref{eq:kuiper-statistic}], whether they were uniformly distributed on the interval $[0,1)$.  Unsurprisingly, this was not the case, as seen in the insets of Fig.~\ref{fig:single_population}.  We therefore proceeded to fit the data with a mixture of $K = 1, 2, \dots, 15$ subpopulations via the EM algorithm of Sec.~\ref{sec:em-algorithm}, starting from multiple different random initial values for $\smash{\{ a_{k}^{2}, \sigma_{k}^{2} \}_{k=1,\dots,K}}$, and uniform mixing fractions $P_{k} = 1/K$ $\forall k$.  This was done to increase the chance of finding the global minimum of Eq.~\eqref{eq:em-log-likelihood}.  Further details on our implementation of the EM algorithm can be found in Appendix~\ref{app:em-algorithm}.  

\begin{figure*}[t!]
\centering
\includegraphics{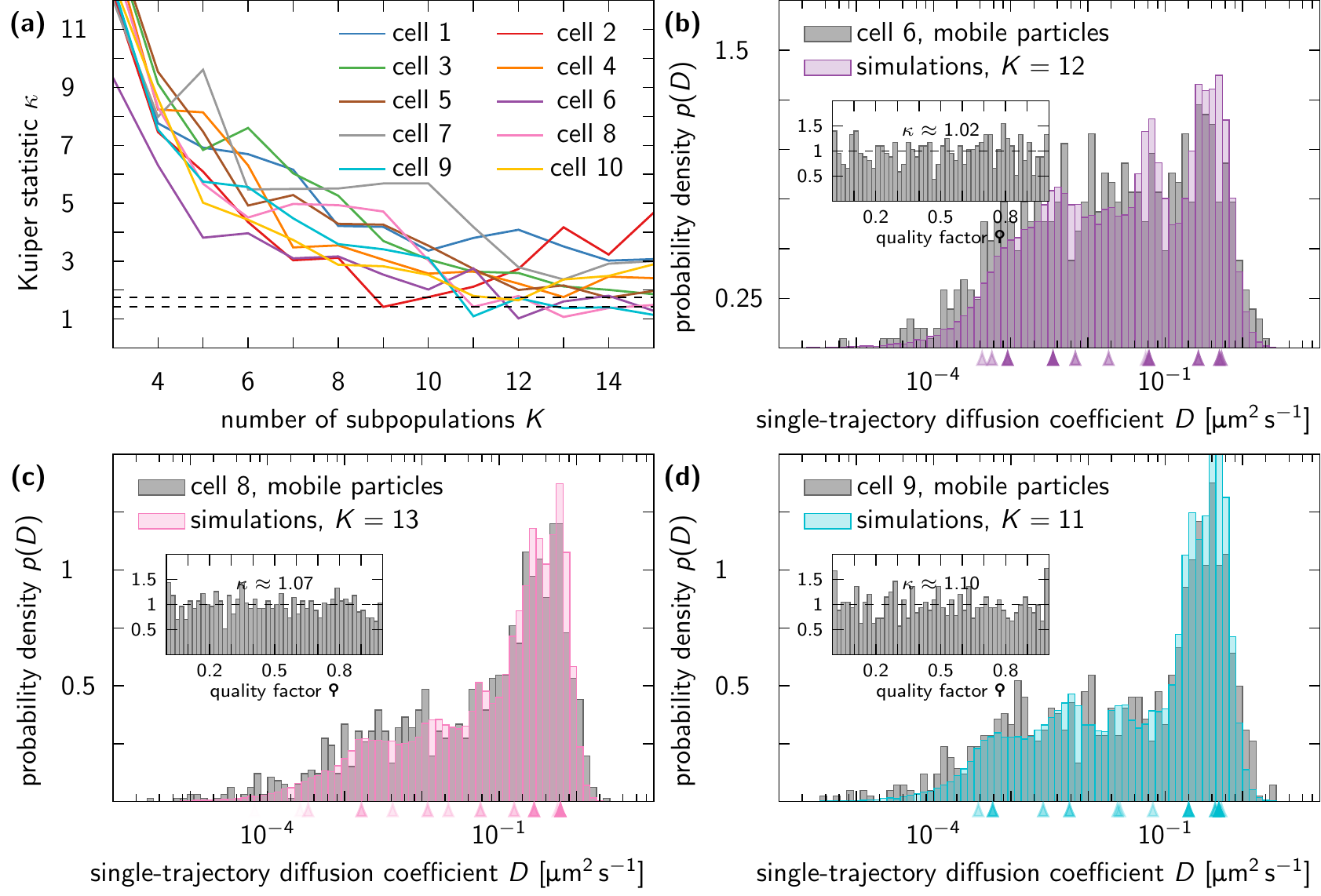}
\caption{Multi-mixture models compared to experimental data.  (a)~The Kuiper statistic [Eq.~\eqref{eq:kuiper-statistic}] for each cell as a function of the number $K$ of subpopulations in the mixture.  The black dashed lines mark the thresholds $\kappa \approx 1.42$ and $\kappa \approx 1.75$, below which there is a $>25\%$ and $>5\%$ chance, respectively, to obtain quality factor results at least as extreme as actually observed.  Three cells clearly cross the lower threshold, and a fourth one narrowly misses it.  (b-d)~Comparison between experimental single-trajectory diffusion coefficient distributions (grey histograms) and results from Brownian dynamics simulations.  The $10^{5}$ synthetic diffusion coefficients making up each of the colored histograms were obtained in almost the same way as in Fig.~\ref{fig:single_population}, except that different parameters $a_{k}^{2}$ and $\sigma_{k}^{2}$ were used for each mixture component, and the mixing fractions $P_{k}$ determined their frequency of occurrence.  The arrows indicate the diffusion coefficients behind the subpopulations and their relative opacities the respective mixing fractions, both of which are tabulated in Table~\ref{tab:diffusion-parameters}.  \emph{Insets:} Distributions of quality factors, after sorting the trajectories according to the subpopulations they most likely belong to.  }
\label{fig:multiple_populations}
\end{figure*}

\begin{table}[t!]
\caption{Fit parameters for specific cells obtained via the EM algorithm.  The number of trajectories $M$, mean trajectory length $\avg{N+1}$ and standard deviation (SD) of the trajectory lengths are listed for each cell.  The units of the diffusion coefficients $D_{k}$ and static noise amplitudes $\smash{a_{k}^{2}}$ are \si{\micro \meter \squared \per \second} and \si{\micro \meter \squared}, respectively.  }
\scriptsize
\begin{center}
\begin{tabularx}{\linewidth}{l X X X}
& \textbf{cell 6} & \textbf{cell 8} & \textbf{cell 9}
\\
& $M = 874$ & $M = 1582$ & $M = 1178$
\\
& $\avg{N+1} = 26$ & $\avg{N+1} = 30$ & $\avg{N+1} = 30$
\\
& $\operatorname{SD} = 44$ & $\operatorname{SD} = 56$ & $\operatorname{SD} = 49$
\\
\toprule
$D_{1}$ & \SI{5.4(4)e-1}{} & \SI{6.2(1)e-1}{} & \SI{5.3(5)e-1}{}
\\
$a_{1}^{2}$ & \SI{2.5(2)e-2}{} & \SI{4.3(3)e-3}{} & \SI{2.8(2)e-2}{}
\\
$P_{1}$ & 0.09 & 0.223 & 0.06
\\
\midrule
$D_{2}$ & \SI{5.0(2)e-1}{} & \SI{5.7(4)e-1}{} & \SI{5.0(2)e-1}{}
\\
$a_{2}^{2}$ & \SI{4.7(5)e-3}{} & \SI{2.0(1)e-2}{} & \SI{7.3(4)e-3}{}
\\
$P_{2}$ & 0.11 & 0.052 & 0.17
\\
\midrule
$D_{3}$ & \SI{2.7(1)e-1}{} & \SI{2.83(7)e-1}{} & \SI{4.4(2)e-1}{}
\\
$a_{3}^{2}$ & \SI{2.0(2)e-3}{} & \SI{3.4(1)e-3}{} & \SI{1.8(3)e-3}{}
\\
$P_{3}$ & 0.11 & 0.184 & 0.12
\\
\midrule
$D_{4}$ & \SI{6.2(4)e-2}{} & \SI{1.55(5)e-1}{} & \SI{2.01(6)e-1}{}
\\
$a_{4}^{2}$ & \SI{7.9(3)e-3}{} & \SI{2.3(1)e-3}{} & \SI{2.6(1)e-3}{}
\\
$P_{4}$ & 0.11 & 0.089 & 0.16
\\
\midrule
$D_{5}$ & \SI{6.1(3)e-2}{} & \SI{5.8(5)e-2}{} & \SI{6.8(3)e-2}{}
\\
$a_{5}^{2}$ & \SI{1.9(1)e-3}{} & \SI{1.42(5)e-2}{} & \SI{2.5(1)e-3}{}
\\
$P_{5}$ & 0.06 & 0.046 & 0.05
\\
\midrule
$D_{6}$ & \SI{5.6(9)e-2}{} & \SI{5.7(2)e-2}{} & \SI{2.6(2)e-2}{}
\\
$a_{6}^{2}$ & \SI{2.3(1)e-2}{} & \SI{1.58(5)e-3}{} & \SI{9.2(3)e-3}{}
\\
$P_{6}$ & 0.03 & 0.065 & 0.04
\\
\midrule
$D_{7}$ & \SI{1.8(1)e-2}{} & \SI{2.2(1)e-2}{} & \SI{2.4(1)e-2}{}
\\
$a_{7}^{2}$ & \SI{1.19(6)e-3}{} & \SI{5.1(1)e-3}{} & \SI{2.02(6)e-3}{}
\\
$P_{7}$ & 0.07 & 0.053 & 0.06
\\
\midrule
$D_{8}$ & \SI{6.9(7)e-3}{} & \SI{1.18(4)e-2}{} & \SI{5.8(3)e-3}{}
\\
$a_{8}^{2}$ & \SI{6.0(2)e-3}{} & \SI{1.51(3)e-3}{} & \SI{1.54(3)e-3}{}
\\
$P_{8}$ & 0.08 & 0.080 & 0.10
\\
\midrule
$D_{9}$ & \SI{3.5(3)e-3}{} & \SI{4.1(3)e-3}{} & \SI{2.6(3)e-3}{}
\\
$a_{9}^{2}$ & \SI{2.15(5)e-3}{} & \SI{3.77(8)e-3}{} & \SI{3.81(9)e-3}{}
\\
$P_{9}$ & 0.13 & 0.045 & 0.07
\\
\midrule
$D_{10}$ & \SI{9.2(7)e-4}{} & \SI{1.64(8)e-3}{} & \SI{5.8(4)e-4}{}
\\
$a_{10}^{2}$ & \SI{1.49(3)e-3}{} & \SI{1.38(2)e-3}{} & \SI{1.34(2)e-3}{}
\\
$P_{10}$ & 0.12 & 0.103 & 0.12
\\
\midrule
$D_{11}$ & \SI{5.6(8)e-4}{} & \SI{3.3(3)e-4}{} & \SI{3.8(3)e-4}{}
\\
$a_{11}^{2}$ & \SI{9.9(3)e-4}{} & \SI{9.2(2)e-4}{} & \SI{6.7(1)e-4}{}
\\
$P_{11}$ & 0.05 & 0.044 & 0.05
\\
\midrule
$D_{12}$ & \SI{4.2(8)e-4}{} & \SI{2.6(5)e-4}{} & 
\\
$a_{12}^{2}$ & \SI{2.90(8)e-3}{} & \SI{5.7(2)e-4}{} & 
\\
$P_{12}$ & 0.04 & 0.012 & 
\\
\midrule
$D_{13}$ &  & \SI{7(5)e-5}{} & 
\\
$a_{13}^{2}$ &  & \SI{3.4(3)e-4}{} & 
\\
$P_{13}$ &  & 0.004 & 
\\
\bottomrule
\end{tabularx}
\end{center}
\label{tab:diffusion-parameters}
\end{table}

Although our results for the live-cell data were not entirely unambiguous, the general trend of $\kappa$ decreasing with increasing $K$ was still observed [see Fig.~\ref{fig:multiple_populations}(a)].  Depending on the choice of threshold, we either ended up with six or three cells, for which the null hypothesis that their single-trajectory diffusion coefficient distributions originated from a finite mixture of fast and slowly diffusing particles could not be rejected.  Furthermore, the three cells that crossed the more stringent threshold of $\kappa \approx 1.42$ all gave $\kappa$-values with associated $p$-values [Eq.~\eqref{eq:p-value}] well above $0.5$.  For cells 6 and 9, the same optimal $K$-values were predicted for both thresholds, whereas the optimum shifted somewhat upwards for cell 8.  

The results of the lower threshold suggest that the trajectories of cells 6, 8 and 9 should be sorted into 11 to 13 distinct subpopulations [see Figs.~\ref{fig:multiple_populations}(b-d)], whose relevant parameters are tabulated in Table~\ref{tab:diffusion-parameters}.  However, it should be noted that some of the populations can be lumped together if we solely focus on the diffusion coefficient.  Also, the last two subpopulations of cell 8 are extremely underrepresented and can, in principle, be neglected.  While this may not be apparent from the corresponding $P_{k}$-values, it becomes clear when the classification coefficients $T_{k,m}$ are inspected: It turns out that for the parameters presented here only 9 and 6 trajectories out of a total of 1349 mobile ones get assigned to the 12th and 13th subpopulations, respectively.  This might be an indication of an overly stringent threshold, and that cell 8 is better described with $K=11$ subpopulations.  In conclusion, the effective number of subpopulations within each cell might therefore be somewhat smaller than the output of the EM algorithm seems to imply, but certainly not two or three.  

\begin{figure}[t!]
\centering
\includegraphics{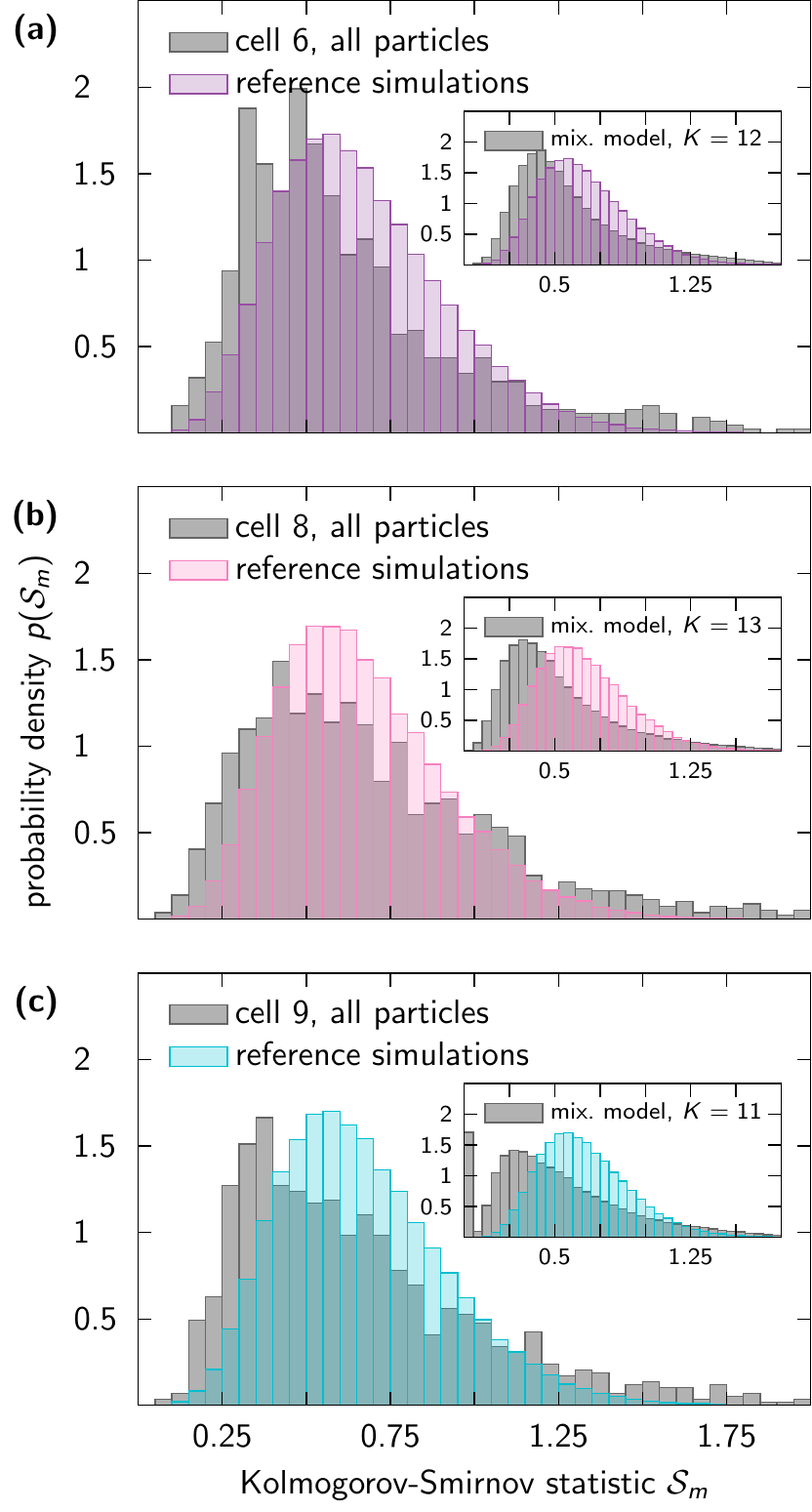}
\caption{Distributions of Kolmogorov-Smirnov statistics for the cells analyzed in Fig.~\ref{fig:multiple_populations}.  Using global single-population estimates for $a^{2}$ and $\sigma^{2}$, we estimated how likely it is for early parts of a trajectory to be governed by the same diffusion dynamics as its later parts via two-sample KS tests.  The results for (a)~cell 6, (b)~cell 8 and (c)~cell 9 (gray histograms) are consistent and show shifted weights toward smaller $\mathcal{S}_{m}$-values and broader tails than the corresponding reference distributions (colored histograms).  \emph{Insets:} Analogous plots for Brownian dynamics simulation data generated using the mixture model parameters behind Figs.~\ref{fig:multiple_populations}(b-d) (see also Table~\ref{tab:diffusion-parameters}).  We considered $10^{5}$ trajectories with lengths distributed in the same way as in the experiments.  }
\label{fig:KS_statistics}
\end{figure}

For completeness, in Fig.~\ref{fig:KS_statistics} we plot the KS distributions for cells 6, 8 and 9, as well as for the multi-population simulation data of Fig.~\ref{fig:multiple_populations}.  In comparison to single-population reference data, we can see that both the cell data and multi-population simulation data are shifted and have a slightly broader tail than the reference.  The qualitative agreement between the cell data and the simulation data strengthens our working hypothesis that the measured trajectories originate from a heterogeneous mixture of fast and slowly diffusing particles.  However, in light of our results in Sec.~\ref{sec:simulation-data}, we cannot exclude the possibility that the experimental data originate from an underlying intricate PDD profile.  

Note that there appears to be a weak negative correlation between the average trajectory length and the diffusion coefficient associated with each subpopulation.  This means that the trajectories of fast diffusing populations are generally shorter than the trajectories of slow diffusing populations.  We can rule out effects due to the trajectory-length bias of the MLEs (see Fig.~\ref{fig:mle_bias}), because said bias is virtually non-existent in global estimates of trajectories of length $N+1 \geq 5$ and the experimental datasets contain exclusively trajectories of length 8 or greater.  The correlation is possibly explained by the fact that tracking algorithms have difficulties stitching together trajectories of fast diffusing particles with large displacements between frames, and are therefore more likely to split them up into shorter trajectories.  

\begin{figure*}[t!]
\centering
\includegraphics{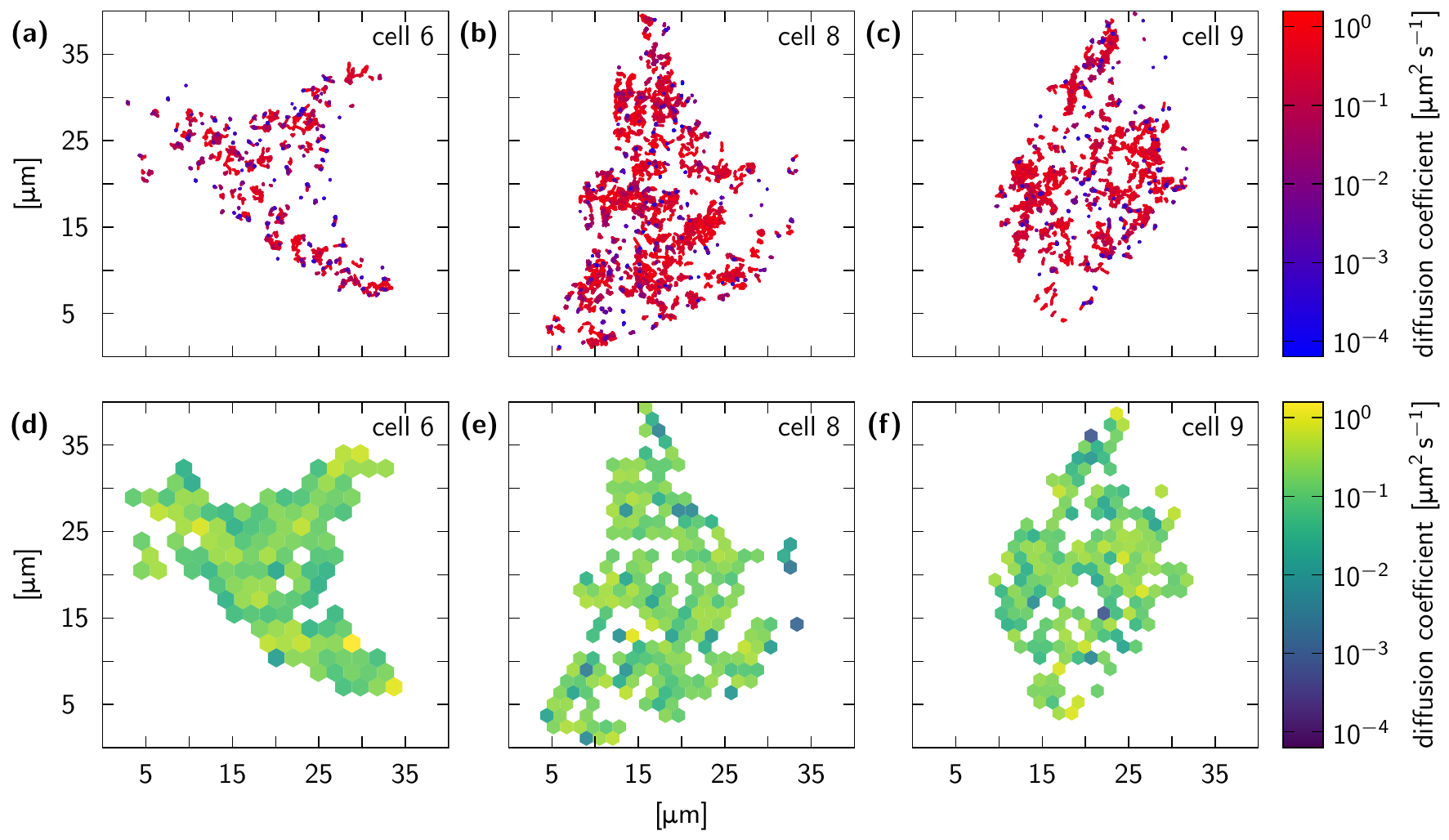}
\caption{Spatial distribution of fast and slowly diffusing MET receptor tyrosine kinases.  Using the classification coefficients [Eq.~\eqref{eq:T-coefficients}], we assign each trajectory to an appropriate subpopulation.  Depending on their diffusion coefficients (see also Table~\ref{tab:diffusion-parameters}), the subpopulations have been color-coded from blue (slow) to red (fast).  As in Fig.~\ref{fig:multiple_populations}, we focus on the three cells whose diffusion coefficient distributions are well described with the mixture model, namely (a)~cell~6, (b)~cell~8, and (c)~cell~9.  (d-f)~Diffusivity landscapes inferred from mobile and immobile particles using the software platform \emph{TRamWAy}.  The localization uncertainty was estimated as $a \approx \SI{0.05}{\micro \meter}$ from the values in Table~\ref{tab:diffusion-parameters}, and the results seemed independent of the value chosen for the regularization factor.  Note that the resolution of the diffusivity landscape is lower than the trajectory representation in (a-c), which also results in the more moderate diffusion coefficient values.  
}
\label{fig:trajectories}
\end{figure*}

Finally, we want to demonstrate how the information encoded in the classification coefficients $T_{k,m}$ can be employed, such as specifically picking out trajectories belonging to a certain subpopulation for further analysis.  Figure~\ref{fig:trajectories}(a-c) visualizes the trajectories of mobile particles in cells 6, 8 and 9.  We indicate increasing receptor diffusivity with a color code from blue to red. The resulting map of confined blue and spread-out red trajectories is reminiscent of the diffusivity landscapes mapped out by Gaussian propagators in Refs.~\onlinecite{El-BeheiryDahan2015} and~\onlinecite{LaurentFloderer2020}.  Motivated by this observation, we applied the \emph{TRamWAy} open-source software platform for analyzing single biomolecule dynamics~\cite{LaurentFloderer2020} to the tracking data of our best-behaving cells.  We thereby relied on the diffusion-only (\texttt{standard.d}) inference mode with the hyperparameter of the diffusivity prior (\texttt{diffusivity\_prior}) and the minimum number of assigned (trans-)locations (\texttt{min\_location\_count}) both set to 1, and a nearest neighbor assignment by count (\texttt{from\_nearest\_neighbors}) of 10.  The localization precision (\texttt{localization\_precision}) was chosen as the geometric average of the $a^{2}$-values reported in Table~\ref{tab:diffusion-parameters}, giving 0.05.  The results are depicted in Figs.~\ref{fig:trajectories}(d-f).  All in all, the inferred diffusivity landscapes do not have the same resolution as Figs.~\ref{fig:trajectories}(a-c), where  immediately adjacent trajectories can have vastly different diffusion coefficients, and therefore do not cover the same range of diffusivity as the mixture model does.  Furthermore, \emph{TRamWAy} does not consider $a^{2}$ as a parameter to infer, but a constant specified by the user. Whereas it could be insightful to analyze the data in Fig.~\ref{fig:trajectories} also in the context of corralled diffusion~\cite{Saxton1995}, we leave a more definitive analysis of the spatial patterning of cell-surface receptor diffusion for future studies.  

\section{Outlook: Towards Bayesian inference}\label{sec:bayesian-inference}

We want to point out that both the single-population log-likelihood [Eq.~\eqref{eq:M-log-likelihood}] and the mixture log-likelihood [Eq.~\eqref{eq:em-log-likelihood}] can be paired with priors $\Pi(\smash{a^{2}},\smash{\sigma^{2}})$ to construct posterior distributions via Bayes' theorem.  This not only affects the numerical values of the parameter estimates, but also their interpretation, because in a Bayesian setting the model parameters are assumed to be randomly distributed.  A Bayesian prior can, \emph{e.g.}, be used to allow for variations in the localization uncertainty $a^{2}$ for each recorded particle individually.  

The priors can either be chosen as (weakly) informative, where definite information about a parameter is taken into consideration, or uninformative.  We expect informative prior information (if any) to exist for $a^{2}$, which can often be estimated empirically, \emph{e.g.}, via auxiliary experiments involving immobilized labeled particles.  Its estimated mean $\smash{a_{\text{emp}}^{2}}$ and uncertainty $\smash{\delta a_{\text{emp}}^{2}}$ can be incorporated into an informative prior, such as
\begin{equation*}
\Pi (a^{2}) \propto \exp \bigg( - \frac{(a^{2} - a_{\text{emp}}^{2})^{2}}{2 (\delta a_{\text{emp}}^{2})^{2}} \bigg) \, ,
\end{equation*}
which then enters the joint prior $\Pi (a^{2},\sigma^{2})$ as follows,
\begin{equation*}
\Pi (a^{2},\sigma^{2}) = \Pi (\sigma^{2} \mid a^{2}) \Pi (a^{2}) \, .  
\end{equation*}
The remaining prior $\Pi (\sigma^{2} \mid a^{2})$ should be chosen as uninformative if no substantial information on $\sigma^{2}$ is available.  The Jeffreys prior~\cite{Jeffreys1946} is a classic uninformative prior, which is proportional to the square root of (the determinant of) the Fisher information and therefore invariant under coordinate transformations.  Using the coordinates $a^{2}$ and $\phi$, the Jeffreys prior for a fixed $a^{2}$ corresponds to the square root of the $I_{2,2}$-element of the matrix $\smash{\mathbf{I}(a^{2},\phi)}$ (see Appendix~\ref{app:fisher-information}), namely
\begin{equation*}
\Pi (\phi \mid a^{2}) = \sqrt{ - \frac{d}{2} \sum_{m=1}^{M} \frac{\mathrm{d}^{2} \ln(\vert \mathbf{\tilde{\Sigma}}_{m} \vert)}{\mathrm{d} \phi^{2}}} \equiv \Pi (\phi) \, ,
\end{equation*}
and is, in fact, independent of $a^{2}$.  Alternatively, one can assume a uniform prior for $\phi$, \emph{i.e.}
\begin{equation*}
\Pi (\phi \mid a^{2}) = \const
\end{equation*}

If no information is available on any of the two parameters, the priors can either be chosen to differ between parameters or be of the same type.  The two-parameter Jeffreys prior is given by
\begin{equation*}
\Pi (a^{2},\phi) = \sqrt{\det \big( \mathbf{I}(a^{2},\phi) \big)} \, ,
\end{equation*}
where an analytic expression for the Fisher information matrix $\mathbf{I}(a^{2},\phi)$ can be found in Appendix~\ref{app:fisher-information}.  While it is usually not recommended to use Jeffreys prior on anything other than single-parameter models, it turns out that for the present problem the preferred alternative, the so-called reference prior~\cite{Bernardo1979, BergerBernardo2009}, coincides with the bivariate Jeffreys prior.  This might be due to the fact that the Jeffreys prior becomes separable when treated in the $(a^{2},\phi)$-coordinates, or because both parameters are scale parameters.  

Bayesian estimates are, in comparison to MLEs, often more sharply distributed and thus have smaller uncertainties.  However, poorly chosen priors introduce biases.  For uninformative priors, the bias vanishes as the sample size goes to infinity, but different priors result in different biases.  Also, the specific choice of the estimator affects the size of the bias and the way it scales with the sample size.  We therefore recommend that practitioners gauge which uninformative priors work best with their estimator of choice.

\section{Conclusions}\label{sec:conclusions}

In this paper, we have derived a numerically efficient maximum likelihood estimator to extract diffusion coefficients from single-molecule tracking experiments subject to static and dynamic noise.  To estimate diffusion coefficients from molecular dynamics simulations, one sets the blur coefficient to $B=0$.  The estimator is based on the statistics of trajectory increments in real space and therefore complements the theory presented in Refs.~\onlinecite{Berglund2010} and~\onlinecite{MichaletBerglund2012}, where an estimator operating in discrete sine space was developed. Fourier representations are computationally efficient and naturally lead, \emph{e.g.}, to a power spectral analysis of the diffusion process~\cite{VestergaardBlainey2014}.  Our approach has the benefit of delivering closed-form expressions for some quantities, such as the accompanying Fisher information matrix.  Compared to other diffusion coefficient estimators, the maximum likelihood estimator has several advantages, \emph{e.g.}, it allows for the inclusion of prior knowledge in a Bayesian manner, makes the error analysis for trajectories of different lengths straightforward, and leads naturally to a quality factor analysis, which helps us to assess whether the data satisfy the underlying assumptions of the diffusion model or not.  

In case of heterogeneous data, the theory can be extended to a mixture model, where the overall numerical efficiency is retained for each subpopulation.  Minimizing the corresponding negative log-likelihood is then achieved with the help of the expectation-maximization algorithm.  To demonstrate its applicability and effectiveness, we first considered sparse quantities of non-diffusive synthetic data to test whether the theory was able to detect any deviations from regular diffusion.  Our results highlight the difficulty to distinguish between diffusive and non-diffusive processes, especially when the trajectories are overwhelmingly short and corrupted by static and dynamic noise.  We then used the framework to analyze single-molecule tracking data for ``resting'' MET receptor tyrosine kinase, recorded in ten different cells~\cite{HarwardtYoung2017}.  As the fit gradually got better with increasing number of subpopulations $K$, we relied on the distribution of quality factors, which are a measure of how well a dataset conforms to the assumptions of the underlying diffusion model, to determine the optimal value of $K$.  Although our results varied significantly between the cells, we found solutions for four of the ten cells, where the null hypothesis that the tracking data originates from a mixture model could not be rejected.  All these fits involved fairly large $K$-values, so we ruled out models containing only a few subpopulations.  However, the effective number of subpopulations can often be reduced, \emph{e.g.}, by neglecting sparsely represented subpopulations or by lumping together subpopulations with similar diffusion coefficients.  Overall qualitative agreement between the experimental data and simulations suggests that the broad distributions of diffusion coefficients observed in the experiment are a result of heterogeneous mixtures, although position-dependent diffusion as an underlying process cannot be ruled out.  In fact, for short trajectories that probe only local regions, position dependence and heterogeneity in diffusion are intertwined.  

We believe that our results provide practitioners of single-molecule tracking techniques with valuable tools to analyze their data.  To facilitate their use, we have implemented the theoretical formulations in a Julia data analysis package~\cite{JuliaScript}.  The general formulation and efficient evaluation of the likelihood functions should also pave the way for more elaborate analysis methods in the future, \emph{e.g.}, involving Bayesian statistics.

\section*{Supplementary Material}
A supplementary file contains detailed information on the trajectory-length distribution for the ten cells analyzed, the model parameters and their uncertainties from fits with different numbers $K$ of subpopulations, and the results of the quality factor analysis in terms of the Kuiper statistic $\kappa$ and the associated \emph{p}-value.

\section*{Data availability}
The data that support the findings of this study are available from the corresponding author upon reasonable request.  Requests for the tracking data should be directed to Prof.~Dr.~Mike Heilemann, corresponding author of the original publication.

\begin{acknowledgments}

We are grateful to Prof.~Dr.~Mike Heilemann, and his current and former lab members, especially Dr.~Sebastian Malkusch, for providing us with the experimental tracking data.  This research was supported by the Max Planck Society and by the LOEWE program (Landes-Offensive zur Entwicklung Wissenschaftlich-\"{o}konomischer Exzellenz) of the state of Hesse conducted within the framework of the MegaSyn Research Cluster.

\end{acknowledgments}

\begin{appendix}

\section{Analytic calculation of determinant}\label{app:recurrence-relation}

Consider an $N \times N$ tridiagonal matrix with main, super and subdiagonal elements $\{ \alpha_{1}, \alpha_{2}, \dots, \alpha_{N} \}$, $\{ \beta_{1}, \beta_{2}, \dots, \beta_{N-1} \}$ and $\{ \gamma_{1}, \gamma_{2}, \dots, \gamma_{N-1} \}$, respectively.  Its determinant can be computed using the following three-term recurrence relation~\cite{El-Mikkawy2004},
\begin{gather}\label{eq:determinant-recurrence-relation}
f_{n} = \alpha_{n} f_{n-1} - \beta_{n-1} \gamma_{n-1} f_{n-2} \, , 
\\
\notag
\begin{aligned}
f_{-1} = 0 \, , & & f_{0} = 1 \, .  
\end{aligned}
\end{gather}
The covariance matrix $\mathbf{\Sigma}$ of the main text [Eq.~\eqref{eq:covariance-matrix}] is not only tridiagonal but also a symmetric Toeplitz matrix, which is why we have $\alpha_{i} = \alpha$ and $\gamma_{i} = \beta_{i} = \beta$ $\forall i$.  Multiplying both sides of Eq.~\eqref{eq:determinant-recurrence-relation} with $z^{n}$ and summing over $n$ results in the generating function
\begin{equation}\label{eq:generating-function}
w(z) = \sum_{n=0}^{\infty} f_{n} z^{n} = \frac{1}{1 - \alpha z + \beta^{2} z^{2}} \, ,
\end{equation}
whose coefficients $f_{n}$ satisfy Eq.~\eqref{eq:determinant-recurrence-relation}.  The denominator of Eq.~\eqref{eq:generating-function} can be factored and the resulting expression rewritten as follows via partial fraction decomposition,
\begin{equation*}
w(z) = \frac{1}{r_{+} - r_{-}} \bigg[ \frac{r_{+}}{1 - r_{+} z} - \frac{r_{-}}{1 - r_{-} z} \bigg] \, ,
\end{equation*}
where $r_{\pm} = (\alpha \pm \sqrt{\alpha^{2} - 4 \beta^{2}}) / 2$.  Each term can be further expanded into a geometric series of $z$ to finally reveal that $f_{n}$ has the functional form
\begin{equation}
f_{n} = \alpha^{n} \frac{(1 + q)^{n+1} - (1 - q)^{n+1}}{2^{n+1} q}
\end{equation}
with $q = \sqrt{1 - 4 \beta^{2} / \alpha^{2}}$.  The determinant of an $N \times N$ symmetric tridiagonal Toeplitz matrix is given by $f_{N}$, which finally results in Eq.~\eqref{eq:general-determinant} of the main text.

\section{Positive definiteness of the covariance matrices}\label{app:positive-definiteness}

An $N \times N$ matrix $\mathbf{M}$ is said to be positive definite if $\vec{x}^{T} \mathbf{M} \vec{x} > 0$ $\forall \vec{x} \in \mathbb{R}^{N} \setminus \vec{0}$, which implies that all of its eigenvalues $\lambda_{n}$ must be strictly positive.  The eigenvalues of a symmetric tridiagonal Toeplitz matrix with diagonal and off-diagonal elements $\alpha, \beta \in \mathbb{R}$, respectively, are given by
\begin{equation*}
\lambda_{n} = \alpha - 2 \vert \beta \vert \cos \bigg( \frac{n \pi}{N+1} \bigg) > \alpha - 2 \vert \beta \vert
\end{equation*}
for $n = 1, 2, \dots, N$.  The covariance matrices are therefore positive definite if they are diagonally dominant, \emph{i.e.}~$\alpha - 2 \vert \beta \vert \geq 0$.  This is obviously the case for $\mathbf{\Sigma'}$ and $\mathbf{\Sigma''}$ [Eqs.~\eqref{eq:sigma-prime} and~\eqref{eq:sigma-double-prime}], because $0 \leq B \leq 1/4$ must hold.  Furthermore, because of
\begin{align*}
1 + \phi (1 & - 2B) - 2 \bigg\vert - \frac{1}{2} + \phi B \bigg\vert
\\
& = 
\begin{cases}
2 + (1 - 4 B) \phi \, , & \phi > (2 B)^{-1}
\\
\phi \, , & \phi \leq (2 B)^{-1}
\end{cases} \, ,
\end{align*}
the matrices $\mathbf{\Sigma}$ and $\mathbf{\tilde{\Sigma}}$ are also positive definite.  

Finally, it should be mentioned that the inverse of a positive definite (nonsingular) symmetric matrix $\mathbf{M}$ is also positive definite, because its eigenvalues are the reciprocals of those of the original matrix.  These results guarantee the positivity of $\smash{\vec{\Delta}^{T} \mathbf{\Sigma'}^{-1} \vec{\Delta}}$,  $\smash{\vec{\Delta}^{T} \mathbf{\Sigma''}^{-1} \vec{\Delta}}$ and $\smash{\vec{\Delta}^{T} \mathbf{\tilde{\Sigma}}^{-1} \vec{\Delta}}$, and the associated parameter estimates.

\section{The Fisher information and error estimates}\label{app:fisher-information}

By definition, the Fisher information matrix
\begin{equation*}
\mathbf{I}(a^{2},\phi) = 
\begin{pmatrix}
I_{1,1} & I_{1,2}
\\
I_{2,1} & I_{2,2}
\end{pmatrix}
\end{equation*}
equals the ensemble-averaged Hessian of $\smash{\mathcal{L}(\{ \vec{\Delta}_{m} \} \mid a^{2}, \phi)}$ [Eq.~\eqref{eq:new-log-likelihood}] with the following elements,
\begin{gather*}
\begin{aligned}
I_{1,1} = \bigg\langle \frac{\partial^{2} \mathcal{L}}{(\partial a^{2})^{2}} \bigg\rangle \, , & & I_{2,2} = \bigg\langle \frac{\partial^{2} \mathcal{L}}{\partial \phi^{2}} \bigg\rangle \, ,
\end{aligned}
\\
I_{1,2} = I_{2,1} = \bigg\langle \frac{\partial^{2} \mathcal{L}}{\partial a^{2} \partial \phi} \bigg\rangle \, .  
\end{gather*}
On the boundary, the Fisher information is a scalar function given by
\begin{equation*}
I(\theta) = \bigg\langle \frac{\partial^{2} \mathcal{L}( \{ \vec{\Delta}_{m} \} \mid \theta)}{\partial \theta^{2}} \bigg\rangle \, ,
\end{equation*}
where $\smash{\mathcal{L}( \{ \vec{\Delta}_{m} \} \mid \theta)}$ is defined as $\smash{\mathcal{L}( \{ \vec{\Delta}_{m} \} \mid a^{2}, \sigma^{2})}$ of Eq.~\eqref{eq:M-log-likelihood} with one of the two parameters set to zero, and $\smash{\theta \in \{ a^{2}, \sigma^{2} \}}$ denotes the remaining non-zero parameter.  

Considering first the scalar case, we have $\mathbf{\Sigma} = \theta \, \mathbf{\Sigma}_{\theta}$ with $\smash{\mathbf{\Sigma}_{a^{2}}} = \mathbf{\Sigma}'$ and $\smash{\mathbf{\Sigma}_{\sigma^{2}}} = \mathbf{\Sigma}''$, and therefore
\begin{equation*}
\frac{\partial^{2} \mathcal{L}}{\partial \theta^{2}} = \frac{1}{\theta^{3}} \sum_{m=1}^{M} \sum_{n=1}^{d} \vec{\Delta}_{m,n}^{T} (\mathbf{\Sigma}_{\theta}^{-1})_{m} \vec{\Delta}_{m,n} - \frac{d \NM}{2 \theta^{2}} \, .  
\end{equation*}
Ensemble averages only affect terms dependent on $\smash{\vec{\Delta}}$, which are in this case all of quadratic form and can therefore be computed using $\avg{\Delta_{i} \Delta_{j}} = \Sigma_{i,j} = \theta \, (\Sigma_{\theta})_{i,j}$.  When evaluating the scalar Fisher information, we encounter a sum of terms of the form
\begin{equation*}
\avg{ \vec{\Delta}^{T} \mathbf{\Sigma}_{\theta}^{-1} \vec{\Delta} } = \sum_{i,j=1}^{N} (\Sigma_{\theta}^{-1})_{i,j} \avg{\Delta_{i} \Delta_{j}} = \theta \sum_{i,j=1}^{N} \delta_{i,j} = \theta N \, ,
\end{equation*}
which simply reduce $I(\theta)$ to
\begin{equation*}
I(\theta) = \frac{1}{\theta^{2}} \sum_{m=1}^{M} \sum_{n=1}^{d} N_{m} - \frac{d \NM}{2 \theta^{2}} = \frac{d \NM}{2 \theta^{2}}
\end{equation*}
and thus give rise to the following lower bound for the standard error $\delta \theta = \smash{\sqrt{\var(\theta)}}$,
\begin{equation*}
\delta \theta \geq \sqrt{\frac{1}{I(\theta)}} = \theta \sqrt{\frac{2}{d \NM}} \, .  
\end{equation*}

The general case makes use of the Hessian $2 \times 2$ matrix, whose components are given by
\begin{align*}
\frac{\partial^{2} \mathcal{L}}{(\partial a^{2})^{2}} & = \frac{1}{(a^{2})^{3}} \sum_{m=1}^{M} \sum_{n=1}^{d} \vec{\Delta}_{m,n}^{T} \mathbf{\tilde{\Sigma}}_{m}^{-1} \vec{\Delta}_{m,n} - \frac{d \NM}{2 (a^{2})^{2}} \, ,
\\
\frac{\partial^{2} \mathcal{L}}{\partial a^{2} \partial \phi} & = \frac{1}{2 (a^{2})^{2}} \sum_{m=1}^{M} \sum_{n=1}^{d} \vec{\Delta}_{m,n}^{T} \mathbf{\tilde{\Sigma}}_{m}^{-1} \mathbf{\Sigma}_{m}'' \mathbf{\tilde{\Sigma}}_{m}^{-1} \vec{\Delta}_{m,n} \, ,
\\
\frac{\partial^{2} \mathcal{L}}{\partial \phi^{2}} & = \frac{1}{a^{2}} \sum_{m=1}^{M} \sum_{n=1}^{d} \vec{\Delta}_{m,n}^{T} \mathbf{\tilde{\Sigma}}_{m}^{-1} \mathbf{\Sigma}_{m}'' \mathbf{\tilde{\Sigma}}_{m}^{-1} \mathbf{\Sigma}_{m}'' \mathbf{\tilde{\Sigma}}_{m}^{-1} \vec{\Delta}_{m,n}
\\
& \mathrel{\phantom{=}} + \frac{d}{2} \sum_{m=1}^{M} \frac{\mathrm{d}^{2} \ln (\vert \mathbf{\tilde{\Sigma}}_{m} \vert)}{\mathrm{d} \phi^{2}} \, .  
\end{align*}
according to the identity $\mathrm{d} \mathbf{A}^{-1} / \mathrm{d} \theta = - \mathbf{A}^{-1} (\mathrm{d} \mathbf{A} / \mathrm{d} \theta) \mathbf{A}^{-1}$.  The first diagonal element of the Fisher matrix is calculated in an analogous way to the scalar Fisher information above, giving
\begin{equation*}
I_{1,1} = \frac{d \NM}{2 (a^{2})^{2}} \, .  
\end{equation*}
For the summands of the non-diagonal elements, we find
\begin{align*}
\avg{\vec{\Delta}^{T} \mathbf{\tilde{\Sigma}}^{-1} \mathbf{\Sigma}'' \mathbf{\tilde{\Sigma}}^{-1} \vec{\Delta}}
& = \sum_{\mathclap{i,j,k,l=1}}^{N} \tilde{\Sigma}_{i,j}^{-1} \Sigma_{j,k}'' \tilde{\Sigma}_{k,l}^{-1} \avg{\Delta_{i} \Delta_{l}}
\\
& = a^{2} \sum_{\mathclap{i,j=1}}^{N} \tilde{\Sigma}_{i,j}^{-1} \Sigma_{j,i}'' = a^{2} \tr (\mathbf{\tilde{\Sigma}}^{-1} \mathbf{\Sigma}'') \, .  
\end{align*}
With the help of $\mathbf{\Sigma}'' = \mathrm{d} \mathbf{\tilde{\Sigma}} / \mathrm{d} \phi$ and Jacobi's formula, which states that
\begin{equation*}
\frac{\mathrm{d} \vert \mathbf{A} \vert}{\mathrm{d}\theta} = \vert \mathbf{A} \vert \tr \bigg( \mathbf{A}^{-1} \frac{\mathrm{d} \mathbf{A}}{\mathrm{d} \theta} \bigg)
\end{equation*}
must hold for an invertible matrix $\mathbf{A}$, we can rewrite the trace as follows,
\begin{equation*}
\tr (\mathbf{\tilde{\Sigma}}^{-1} \mathbf{\Sigma}'') = \tr \bigg( \mathbf{\tilde{\Sigma}}^{-1} \frac{\mathrm{d}\mathbf{\tilde{\Sigma}}}{\mathrm{d}\phi} \bigg) = \frac{1}{\vert \mathbf{\tilde{\Sigma}} \vert} \frac{\mathrm{d} \vert \mathbf{\tilde{\Sigma}} \vert}{\mathrm{d}\phi} = \frac{\mathrm{d}\ln (\vert \mathbf{\tilde{\Sigma}} \vert)}{\mathrm{d}\phi} \, .  
\end{equation*}
We therefore get
\begin{equation*}
I_{1,2} = I_{2,1} = \frac{d}{2 a^{2}} \sum_{m=1}^{M} \frac{\mathrm{d}\ln (\vert \mathbf{\tilde{\Sigma}}_{m} \vert)}{\mathrm{d}\phi} \, .  
\end{equation*}
The last diagonal element of the Fisher matrix can be computed with the help of
\begin{align*}
\avg{ \vec{\Delta}^{T} \mathbf{\tilde{\Sigma}}^{-1} & \mathbf{\Sigma}'' \mathbf{\tilde{\Sigma}}^{-1} \mathbf{\Sigma}'' \mathbf{\tilde{\Sigma}}^{-1} \vec{\Delta} }
\\
& = \sum_{\mathclap{i,j,k,l,m,n=1}}^{N} \tilde{\Sigma}_{i,j}^{-1} \Sigma_{j,k}'' \tilde{\Sigma}_{k,l}^{-1} \Sigma_{l,m}'' \tilde{\Sigma}_{m,n}^{-1} \avg{\Delta_{i} \Delta_{n}}
\\
& = a^{2} \sum_{\mathclap{i,j,k,l=1}}^{N} \tilde{\Sigma}_{i,j}^{-1} \Sigma_{j,k}'' \tilde{\Sigma}_{k,l}^{-1} \Sigma_{l,i}''
\\
& = a^{2} \tr ( \mathbf{\tilde{\Sigma}}^{-1} \mathbf{\Sigma}'' \mathbf{\tilde{\Sigma}}^{-1} \mathbf{\Sigma}'' ) \, ,
\end{align*}
where the identity $\mathrm{d} \tr(\mathbf{A}) / \mathrm{d} \theta = \tr (\mathrm{d} \mathbf{A} / \mathrm{d} \theta)$ can be used to rewrite the trace in terms of derivatives, giving
\begin{align*}
\tr \bigg( \mathbf{\tilde{\Sigma}}^{-1} \frac{\mathrm{d} \mathbf{\tilde{\Sigma}}}{\mathrm{d} \phi} \mathbf{\tilde{\Sigma}}^{-1} \frac{\mathrm{d} \mathbf{\tilde{\Sigma}}}{\mathrm{d} \phi} \bigg)
& = - \frac{\mathrm{d}}{\mathrm{d} \phi} \tr \bigg( \mathbf{\tilde{\Sigma}}^{-1} \frac{\mathrm{d} \mathbf{\tilde{\Sigma}}}{\mathrm{d} \phi} \bigg)
\\
& = - \frac{\mathrm{d}^{2} \ln (\vert \mathbf{\tilde{\Sigma}} \vert)}{\mathrm{d}\phi^{2}} \, .  
\end{align*}
The element $I_{2,2}$ thus takes the form
\begin{equation*}
I_{2,2} = - \frac{d}{2} \sum_{m=1}^{M} \frac{\mathrm{d}^{2} \ln (\vert \mathbf{\tilde{\Sigma}}_{m} \vert)}{\mathrm{d}\phi^{2}} \, .  
\end{equation*}
The derivatives of $\ln (\vert \mathbf{\tilde{\Sigma}}_{m} \vert)$ read
\begin{align*}
\frac{\mathrm{d} \ln (\vert \mathbf{\tilde{\Sigma}}_{m} \vert)}{\mathrm{d} \phi} & = \frac{N_{m}}{\tilde{\alpha}} (1 - 2 B_{m}) - F(\tilde{q}) \, , 
\\
\frac{\mathrm{d}^{2} \ln (\vert \mathbf{\tilde{\Sigma}}_{m} \vert)}{\mathrm{d} \phi^{2}} & = - \frac{N_{m}}{\tilde{\alpha}^{2}} (1 - 2B_{m})^{2} - \frac{\mathrm{d} F(\tilde{q})}{\mathrm{d} \phi} \, , 
\end{align*}
where $\tilde{q} = \smash{(1 - 4 \tilde{\beta}^{2} / \tilde{\alpha}^{2})^{1/2}} \geq 0$, $\tilde{\alpha} = 1 + \phi (1 - 2 B_{m})$ and $\tilde{\beta} = - 1 / 2 + \phi B_{m}$.  Here, we introduced the function
\begin{align*}
F(\tilde{q})
& = \frac{2 \tilde{\beta} \big( N_{m} \tilde{q} - 1 + (N_{m}+1) [ \tilde{\alpha} (1 - \tilde{q}) / 2 ]^{N_{m}} / \vert \mathbf{\tilde{\Sigma}}_{m} \vert \big)}{\tilde{\alpha}^{3} \tilde{q}^{2} (1 + \tilde{q})} \, ,
\\
F(0) & = \lim_{\tilde{q} \to 0} F(\tilde{q}) = \frac{2 \tilde{\beta} N_{m} (N_{m}-1)}{3 \tilde{\alpha}^{3}} \, ,
\end{align*}
and its gradient
\begin{gather*}
\frac{\mathrm{d} F(\tilde{q})}{\mathrm{d} \phi} = - \frac{2}{\phi} \bigg[ F(\tilde{q}) - C_{1} F(\tilde{q}) + C_{2} F(0) - \frac{\phi}{2} F(\tilde{q})^{2} \bigg] \, ,
\\
\begin{aligned}
C_{1} = \frac{1}{4 \tilde{\beta}} + \frac{3}{2 \tilde{\alpha}} + \frac{\tilde{\beta} \phi}{\tilde{\alpha}^{3}} \bigg[ \frac{3}{\tilde{q}^{2}} + \frac{\tilde{\alpha}^{2} N_{m}}{2 \tilde{\beta}^{2}} \bigg] \, , & & C_{2} = \frac{3 \phi}{4 \tilde{\alpha} \tilde{\beta} \tilde{q}^{2}} \, ,
\end{aligned}
\end{gather*}
to condense our expressions.  

In the general case, the standard errors emerge from the diagonal elements of the inverse Fisher information matrix $\smash{\mathbf{I}^{-1}(a^{2},\sigma^{2})}$.  This inverse is related to $\smash{\mathbf{I}^{-1}(a^{2},\phi)}$, which is simply given by
\begin{align*}
\mathbf{I}^{-1}(a^{2},\phi)
& = \frac{1}{I_{1,1} I_{2,2} - I_{1,2}^{2}} \begin{pmatrix}
I_{2,2} & - I_{1,2}
\\
- I_{1,2} & I_{1,1}
\end{pmatrix} \, .  
\end{align*}
A coordinate transformation then gives us
\begin{equation*}
\mathbf{I}^{-1}(a^{2},\sigma^{2}) = \mathbf{J}^{-1} \mathbf{I}^{-1}(a^{2},\phi) (\mathbf{J}^{-1})^{T} \vert_{\phi = \sigma^{2} / a^{2}}
\end{equation*}
with an inverse Jacobian
\begin{equation*}
\mathbf{J}^{-1} = \begin{pmatrix}
{\displaystyle \frac{\mathstrut \partial a^{2}}{\mathstrut \partial a^{2}} } & {\displaystyle \frac{\mathstrut \partial a^{2}}{\mathstrut \partial \sigma^{2}} }
\\
{\displaystyle \frac{\mathstrut \partial \phi}{\mathstrut \partial a^{2}} } & {\displaystyle \frac{\mathstrut \partial \phi}{\mathstrut \partial \sigma^{2}} }
\end{pmatrix}^{\mathclap{-1}} = \begin{pmatrix}
1 & 0
\\
\phi & a^{2}
\end{pmatrix} \, .  
\end{equation*}
This gives rise to Eqs.~\eqref{eqs:errors} of the main text.

\section{Numerics}\label{app:numerics}

\subsection{Brownian dynamics simulations}\label{app:brownian-dynamics-simulations}

Interpreting Eqs.~\eqref{eq:Y-process}--\eqref{eq:X-process} in the sense of It\^{o}, we discretize them as follows for $\Delta t = N \delta t$,
\begin{align}\label{eqs:discretized-XYZ}
\begin{split}
Y_{i+1} & = Y_{i} + \frac{\sigma}{\sqrt{N}} R'_{i} \, ,
\\ 
Z_{i} & = \sum_{n=1}^{N} \delta t \, s_{n} Y_{n + i N} \, ,
\\
X_{i} & = Z_{i} + \frac{a}{\sqrt{2}} R_{i} \, .  
\end{split}
\end{align}
Here, $R_{i}$ and $R'_{i}$ denote independent, normal distributed random variables with zero mean and unit variance, and the $s_{n}$ are normalized, \emph{i.e.}
\begin{equation*}
\sum_{n=1}^{N} \delta t \, s_{n} = 1 \, .  
\end{equation*}
In case of position-dependent diffusion, the resulting multiplicative noise produces a spurious drift term~\cite{van-Kampen1988}, so the first line in Eqs.~\eqref{eqs:discretized-XYZ} must be replaced with
\begin{equation*}
Y_{i+1} = Y_{i} + \frac{\sigma(Y_{i}) \sigma'(Y_{i})}{N} + \frac{\sigma(Y_{i})}{\sqrt{N}} R'_{i} \, .  
\end{equation*}
This is the discretized version of Eq.~\eqref{eq:spatially-varying-diffusion} of the main text.  

In this paper, we exclusively chose $N=100$ and $s_{n} = \smash{(N \delta t)^{-1}}$.  All two-dimensional trajectories were generated from two one-dimensional trajectories of equal length under the assumption of isotropic diffusion.

\subsection{Implementation of EM algorithm}\label{app:em-algorithm}

Starting from $\smash{P_{k}^{(0)}} = 1/K$ $\forall k$ and a set $\smash{\{ {\smash{a_{k}^{2}}}^{(0)}, {\smash{\sigma_{k}^{2}}}^{(0)} \}_{k=1,\dots,K}}$ of initial parameter values, chosen randomly on a logarithmic scale, the classification coefficients $\smash{T_{k,m}^{(0)}}$ are initialized according to Eq.~\eqref{eq:T-coefficients}.  We then proceed to update the parameters via Eq.~\eqref{eq:mixing-fractions} and either Eq.~\eqref{eq:em-boundary-a2-solution}, \eqref{eq:em-boundary-sigma2-solution} or Eqs.~\eqref{eq:em-general-a2-solution} and~\eqref{eq:em-general-sigma2-solution}, depending on which solution results in the lowest negative log-likelihood value [Eq.~\eqref{eq:em-log-likelihood}]
\begin{equation*}
L^{(i)} = L (\{ \vec{\Delta}_{m,n} \} \mid \{ P_{k}^{(i)}, {\smash{a_{k}^{2}}}^{(i)}, {\smash{\sigma_{k}^{2}}}^{(i)} \}) \, .  
\end{equation*}
The two steps of the EM algorithm are repeated for a fixed number $N_{\text{local}}$ of iteration steps, or broken off prematurely if the following inequality is satisfied for the tolerance threshold $\varepsilon$,
\begin{equation*}
\frac{L^{(i)} - L^{(i-1)}}{\NM} < \varepsilon \, .  
\end{equation*}
The corresponding solution is saved in the ``high score'', and the algorithm is reinitialized with a new set of initial parameters.  The ``high score'' solution is replaced by the new iteration if the latter results in a lower negative log-likelihood.  The EM algorithm is run in total for $N_{\text{global}}$ times.  

We used $N_{\text{local}} = 500$ and $N_{\text{global}} = 50$ with a tolerance threshold of $\varepsilon = 10^{-10}$ when analyzing the data in Figs.~\ref{fig:proof_of_principle}, \ref{fig:inhomogeneous_diffusion} and~\ref{fig:multiple_populations}.

\end{appendix}

\end{document}